\documentclass{aastex63}

\shorttitle{Inner Cavity of Sz 84}
\shortauthors{Hashimoto et al.}

\graphicspath{{./}{figures/}}

\begin{document}

\title{ALMA Observations of the
Inner Cavity in the Protoplanetary Disk around Sz~84}

\correspondingauthor{Jun Hashimoto}
\email{jun.hashimto@nao.ac.jp}

\author[0000-0002-3053-3575]{Jun Hashimoto}
\affil{Astrobiology Center, National Institutes of Natural Sciences, 2-21-1 Osawa, Mitaka, Tokyo 181-8588, Japan}
\affil{Subaru Telescope, National Astronomical Observatory of Japan, Mitaka, Tokyo 181-8588, Japan}
\affil{Department of Astronomy, School of Science, Graduate University for Advanced Studies (SOKENDAI), Mitaka, Tokyo 181-8588, Japan}

\author{Takayuki Muto}
\affil{Division of Liberal Arts, Kogakuin University, 1-24-2, Nishi-Shinjuku, Shinjuku-ku, Tokyo, 163-8677, Japan}

\author[0000-0001-9290-7846]{Ruobing Dong}
\affil{Department of Physics \& Astronomy, University of Victoria, Victoria, BC, V8P 1A1, Canada}

\author{Yasuhiro Hasegawa}
\affil{Jet Propulsion Laboratory, California Institute of Technology, Pasadena, CA 91109, USA}

\author[0000-0003-2458-9756]{Nienke van der Marel}
\affil{Department of Physics \& Astronomy, University of Victoria, Victoria, BC, V8P 1A1, Canada}

\author[0000-0002-6510-0681]{Motohide Tamura}
\affil{Department of Astronomy, University of Tokyo, 7-3-1 Hongo, Bunkyo-ku, Tokyo 113-0033, Japan}
\affil{Astrobiology Center, National Institutes of Natural Sciences, 2-21-1 Osawa, Mitaka, Tokyo 181-8588, Japan}
\affil{Subaru Telescope, National Astronomical Observatory of Japan, Mitaka, Tokyo 181-8588, Japan}

\author[0000-0001-9248-7546]{Michihiro Takami}
\affil{Institute of Astronomy and Astrophysics, Academia Sinica, 11F of Astronomy-Mathematics Building, AS/NTU No.1, Sec. 4, Roosevelt Rd, Taipei 10617, Taiwan, R.O.C.}

\author[0000-0002-3001-0897]{Munetake Momose}
\affil{College of Science, Ibaraki University, 2-1-1, Bunkyo, Mito, Ibaraki 310-8512, Japan}

\begin{abstract}
We present Atacama Large Millimeter/submillimeter Array (ALMA) observations of a protoplanetary disk around the T Tauri star Sz~84 and analyses of the structures of the inner cavity in the central region of the dust disk. Sz~84's spectral energy distribution (SED) has been known to exhibit negligible infrared excess at $\lambda \lesssim$10~$\mu$m due to the disk's cavity structure. 
Analyses of the observed visibilities of dust continuum at 1.3~mm and the SED indicate that the size of the cavity in the disk of large (millimeter size) dust grains is 8~au in radius and that in the disk of small (sub-micron size) dust grains is 60~au in radius. Furthermore, from the SED analyses, we estimate that the upper limit mass of small dust grains at $r<$60~au is less than $\sim$10$^{-3}$~$M_{\earth}$, which is $\lesssim$0.01~\% of the total (small~$+$~large) dust mass at $r<$60~au. These results suggest that large dust grains are dominant at $r<$60~au, implying that dust grains efficiently grow with less efficient fragmentation in this region, potentially due to weak turbulence and/or stickier dust grains. The balance of grain growth and dust fragmentation is an important factor for determining the size of large dust grains in protoplanetary disks, and thus Sz~84 could serve as a good testbed for investigations of grain growth in such disks.

\end{abstract}


\section{Introduction} \label{sec:intro}

Planet formation is believed to take place in protoplanetary disks. In its earliest stages, dust grains grow. For small (sub-micron size) dust grains, which are comparable to typical interstellar medium dust in size, van der Waals forces cause the dust grains, which exhibit Brownian motion, to stick when they collide. Beyond the millimeter size, dust grains can grow into planetesimals a kilometer (or larger) in size via either direct collisional aggregation of dust grains \citep[e.g.,][]{weid1993,kata2013a} or some instabilities such as streaming instabilities \citep[e.g.,][]{joha2014a}. The accretion of such planetesimals eventually leads to the formation of planets.

Grain growth in protoplanetary disks has been extensively investigated \citep[e.g.,][]{test2014}. Theoretically, grain growth is limited by two processes, namely radial drift and fragmentation. Radial drift occurs when dust grains become large enough to decouple from the disk gas and then experience a head wind \citep{weid77}. Accordingly, dust size is limited when dust grains undergo radial drift. Fragmentation also limits grain growth. It occurs when the relative velocity of colliding dust grains becomes high enough to result in fragmentation rather than a merger.

Observational investigations of grain growth have been conducted at (sub-)millimeter wavelengths \citep[e.g.,][]{test2003,pere2012,test2014}. Optical and infrared (IR) observations can trace the surface layer of the disk due to the layer's large optical depth, whereas (sub-)millimeter observations probe thermal emissions from the disk midplane region, where grain growth is the most efficient. This region becomes
optically thin (or thinner) at longer wavelengths. As dust grains grow, the dust opacity ($\kappa_{\lambda}$), which is approximated as $\kappa_{\lambda} \propto \lambda^{\beta}$, and its index ($\beta$) both decrease \citep[e.g.,][]{miya1993,drai2006}. Based on the Rayleigh-Jeans limit and the optically thin assumption, the observed flux can be expressed as $F_{\lambda} \propto \lambda^{\beta+2}$, and thus multiple wavelength observations can be used to measure $\beta$ even if the absolute dust opacity is unknown.

In addition to determining $\beta$, comparisons of disk structures observed at near-infrared (NIR) and (sub-)millimeter wavelengths are useful for exploring grain growth. When a cavity structure in the central region of a disk (such disks are called transitional disks\footnote{There are two types of transitional disk in the simple classical picture based on SED analyses \citep[e.g.,][]{espaillat+14}. One has a cavity in which the central region of the disk is optically thin, and the other is an optically thick disk separated into inner and outer disks by an optically thin gap. Recent studies on transitional disks by \citet{fran2020a} suggest that the existence of the inner disk responsible for the NIR excess does not correlate with that in the dust continuum at (sub-)millimeter wavelengths, and that this discrepancy may not exist for large dust grains.}; e.g., \citealp{espaillat+14}) is observed directly, it may have different sizes at different wavelengths \citep[e.g.,][]{dong12cavity,garu2013,vandermarel13,vill2019a}. One explanation for the different cavity sizes at different wavelengths is the trapping of large (millimeter size) dust grains \citep[e.g.,][]{rice06} by planet--disk interactions \citep[e.g.,][]{zhu12}. A planet embedded in a disk reduces the surface density of the gas and creates a gas gap at the planetary orbital radius. The gap thereby produces a gas pressure bump at its outer edge where large dust grains are trapped. Because the small dust grains coupled to the gas can flow inside the gas pressure bump, their cavity sizes are smaller. This spatial segregation can be used as a proxy of grain growth in the disk.

Dust growth itself can also spatially segregate cavities without dust trapping. The trend of such segregation is opposite to that mentioned above. Numerical simulations of grain growth \citep[e.g.,][]{dull2005a,birn2012a} show that small dust grains, which are the main components of NIR excesses in the spectral energy distribution (SED), grow into large (millimeter size) dust grains within a time scale of $\lesssim$0.1~Myr at $r=$1~au if dust fragmentation is negligible. When small dust grains in the central region of a protoplanetary disk are removed, the corresponding SED shows a deficit of flux at NIR wavelengths. Because grain growth proceeds in an inside-out manner, which is due to higher gas density and a faster dynamical time scale in the inner disk \citep[e.g.,][]{dull2005a,brauer+2008}, the small and large dust grains are predicted to be preferentially distributed in the outer and inner regions of the disk, respectively. Such opposite distributions of small and large dust grains were reported for the transitional disk around DM~Tau (\citealp{kudo2018}; Hashimoto et~al. submitted), in which the cavity in the disk of small dust grains is inferred to be located at $r=$3~au. Large dust grains are present even within this cavity. These results suggest that small dust grains at a~few~au around DM~Tau efficiently grow into large dust grains with less efficient fragmentation, possibly due to weak turbulence and/or stickier dust grains \cite[e.g.,][]{stei2019}.

In this paper, we report another example of an opposite spatial segregation for a cavity in the disk around the T Tauri star Sz~84 (spectral type: M5, \citealp{alca2014}; $T_{\rm eff}$: 3162~K, \citealp{hend2020a}; $M_{\rm *}$: 0.2~$M_{\odot}$, \citealp{alca2014}; age: 1~Myr, \citealp{alca2014}; distance: 152.6~pc, \citealp{gaia18}; disk inclination: 75\degr, \citealp{ansd2016a}). The SED of Sz~84 shows no or very weak IR excesses at $\lambda \lesssim$10~$\mu$m, which has been interpreted as indicating the presence of an almost clean cavity in the disk of small dust grains at $r\sim$55~au (i.e., negligible small dust grains inside the cavity; e.g., \citealp{meri2010a}). Subsequent Atacama Large Millimeter/submillimeter Array (ALMA) observations of Sz~84 with a beam size of 0$\farcs$3 (46~au) showed no cavity structures in the dust continuum image \citep{ansd2016a}, even though visibility analyses suggested a cavity structure at $r=$20--41~au \citep{tazz2017a,vandermarel18a}. These results suggest that the size of the cavity in the disk of small dust grains ($r\sim$55~au) could be larger than that in the disk of large dust grains ($r=$20--41~au). The $^{13}$CO~$J=3\rightarrow2$ gas disk around Sz~84 reveals no cavity structure in integrated CO visibilities \citep{vandermarel18a}. However, gas cavities tend to be 1.5--3 times smaller than dust cavities \citep{vandermarel+16,vandermarel18a}, and thus the available data may simply not resolve a gas cavity. Furthermore, it is reasonable to assume that the mass accretion rate of Sz~84, $\dot{M}$~$\sim$ 1~$\times $ 10$^{-9}$~$M_{\odot}$/yr \citep{mana14}, is comparable to that of typical T~Tauri~stars \citep{naji15}. These two observational results predict that small dust grains well coupled to the gas exist in the vicinity of the central star (they arrive there by flowing inside the cavity), possibly explaining the smaller size of the cavity in the disk of small dust grains. However, as mentioned above, analyses of the visibilities of the dust continuum and the SED in the literature show the opposite observational results. The origin of Sz~84's cavity in the disk of small dust grains with a moderate mass accretion rate is still unclear \citep[e.g.,][]{mana14}. 

\section{Observations} \label{sec:obs}

ALMA observations of Sz~84 were carried out with band~6 in cycle~3 (ID: 2015.1.01301.S; PI: J.~Hashimoto); they are summarized in Table~\ref{tab:table1}. The data were calibrated using the Common Astronomy Software Applications (CASA) package \citep{mcmu07} with the calibration scripts provided by ALMA. We conducted a self-calibration of the visibilities. The phases were self-calibrated once with fairly long solution intervals (solint=`inf') that combined all spectral windows. The proper motions of Sz~84 were calculated with the function \verb#EPOCH_PROP# in GAIA ADQL (Astronomical Data Query Language\footnote{{\sf https://gea.esac.esa.int/archive/}}). The phase centers were corrected using \verb#fixvis# in the CASA tools. The dust continuum image at band~6 was synthesized by CASA with the \verb#CLEAN# task using a multi-frequency deconvolution algorithm \citep{rau11}. In the \verb#CLEAN# task, we set the $uv$-taper to obtain a nearly circular beam pattern (Table~\ref{tab:table1}). The synthesized dust continuum image is shown in Figure~\ref{fig1}. The root-mean-square (r.m.s.) noise in the region far from the object is 61.3~$\mu$Jy/beam with a beam size of 198~$\times$~195~mas at a position angle (PA) of $-$88.5$\degr$. 

In addition to our own data, two archival dust continuum datasets were used for our visibility analysis in \S~\ref{subsec:mcmc}. They were taken with band~3 in cycle~4 (ID: 2016.1.00571.S; PI: M.~Tazzari) and band~7 in cycle~2 (ID: 2013.1.00220.S; PI: J.~Williams). The calibration procedures described above were applied to these data. These additional observations are summarized in Table~\ref{tab:band3-7} in the Appendix. Because the data quality (spatial resolution and sensitivity) of our band~6 data is better than that of the archival data, whose spatial resolution is $\sim$0\farcs3, we only show the band~6 image in Figure~\ref{fig1}. 

The $^{12}$CO~$J=2\rightarrow1$ line data (Table~\ref{tab:table1}) were extracted by subtracting the continuum in the $uv$ plane using the \verb#uvcontsub# task in the CASA tools. A line image cube with channel widths of 1.0~km/s was produced by the \verb#CLEAN# task. The integrated line flux map (moment~0) and the intensity-weighted velocity map (moment~1) are shown in Figures~\ref{fig1} and \ref{fig4}, respectively. The channel maps at $-$1.0 to $+$11.0~km/s are shown in Figure~\ref{figA1} in the Appendix. The r.m.s.\ noise in the moment~0 map is 8.7~mJy/beam$\cdot$km/s with a beam size of 235~$\times$~161~mas at a PA of $-$2.0$\degr$ and that in the moment~1 map at the 1.0~km/s bin is 2.37~mJy/beam. The peak signal-to-noise ratio (SNR) is 17.9 in the channel map of $+$7.0~km/s. $^{13}$CO~$J=2\rightarrow1$ line emissions were detected with a peak SNR of less than 5, and thus the images are not shown. We did not detect C$^{18}$O~$J=2\rightarrow1$ line emissions.

\begin{deluxetable*}{lcc}
\tabletypesize{\footnotesize}
\tablewidth{0pt} 
\tablenum{1}
\tablecaption{ALMA observations and imaging parameters\label{tab:table1}}
\tablehead{
\colhead{Observations} & \colhead{} & \colhead{} 
}
\startdata
Observing date (UT)         & \multicolumn{2}{c}{2016.Sep.17}                     \\
Project code                & \multicolumn{2}{c}{2015.1.01301.S}                  \\
Time on source (min)        & \multicolumn{2}{c}{29.7}                            \\
Number of antennas          & \multicolumn{2}{c}{38}                              \\
Baseline lengths            & \multicolumn{2}{c}{15.1~m to 2.5~km}               \\
Baseband Freqs. (GHz)       & \multicolumn{2}{c}{219.5, 220.4, 230.5, 232.5}      \\
Channel width   (MHz)       & \multicolumn{2}{c}{0.122, 0.122, 0.122, 15.6}      \\
Continuum band width (GHz)  & \multicolumn{2}{c}{2.3}                            \\
Bandpass calibrator         & \multicolumn{2}{c}{J1517$-$2422}                    \\
Flux calibrator             & \multicolumn{2}{c}{J1427$-$4206}                   \\
Phase calibrator            & \multicolumn{2}{c}{J1610$-$3958}                   \\
New phase center with GAIA  & \multicolumn{2}{c}{15h58m2.50261s, -37d36m3.1145s} \\
\hline \hline
\multicolumn{1}{c}{Imaging} & \multicolumn{1}{c}{Dust continuum}            & \multicolumn{1}{c}{$^{12}$CO~$J=2\rightarrow1$} \\
\hline
Bobust clean parameter      & \multicolumn{1}{c}{$-2.0$}                              & 2.0\\
Outer $uv$-taper parameter  & \multicolumn{1}{c}{50.0~$\times$~0.6~M$\lambda$ with 183.0$\degr$} & \nodata \\
Beam shape                  & \multicolumn{1}{c}{198~$\times$~195~mas at PA of $-$88.5$\degr$}   & 235~$\times$~161~mas at PA of $-$2.0$\degr$\\
r.m.s. noise ($\mu$Jy/beam) & \multicolumn{1}{c}{61.3}                                & 2.37~$\times$~10$^{3}$ at 1.0~km/s bin
\enddata
\end{deluxetable*}

\section{Results} \label{sec:result}

\begin{figure}[ht!]
\begin{centering}
\includegraphics[clip,width=\linewidth]{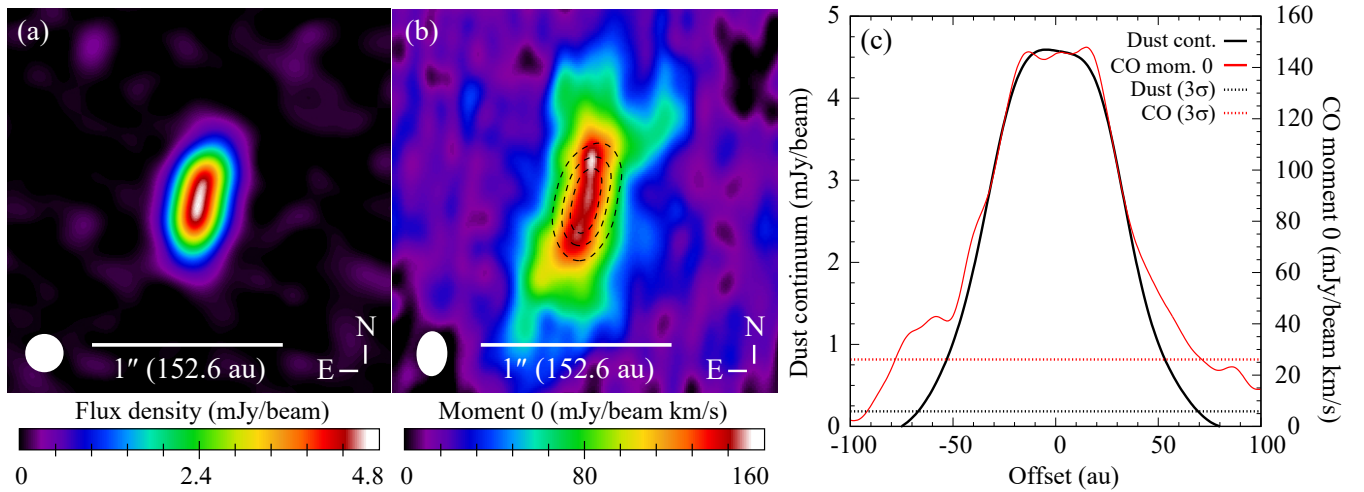}
\end{centering}
\caption{
Observational results for the Sz~84 disk from ALMA at band 6 in cycle 3. 
(a) Dust continuum image. The r.m.s.\ noise is 61.3~$\mu$Jy/beam with a beam size of 198~$\times$~195~mas at a PA of $-$88.5\degr. 
(b) $^{12}$CO moment~0 map. The r.m.s.\ noise is 8.7~mJy/beam$\cdot$km/s with a beam size of 235~$\times$~161~mas at a PA of $-$2.0$\degr$. The contours represent the dust continuum at 20, 40, and 60~$\sigma$. 
(c) Radial cuts in panels (a) and (b) at a PA of 165.71$\degr$ taken from our modeling in \S~\ref{subsec:mcmc}.
} \label{fig1}
\end{figure}

Figures~\ref{fig1}(a) and (b) show the dust continuum image at band~6 and the $^{12}$CO moment~0 map, respectively. Our observations 
clearly detect the spatially resolved disk in both images. However, no cavity structure appears in the central region of the disk in either image. The total flux density of the dust continuum derived from the visibility fitting in \S~\ref{subsec:mcmc} is 13.43~$\pm$~0.13~mJy, which is consistent with previous ALMA band~6 observations (12.63~$\pm$~0.21; \citealp{ansd2018a}) assuming a 10 \% uncertainty in absolute flux calibration. The $^{12}$CO integrated flux density at more than 3~$\sigma$ is 1.89~$\pm$~0.05~Jy/beam$\cdot$km/s.

We derived the spectral index $\alpha$ based on band 3, 6, and 7 data. The total flux densities at bands~3 and 6 taken from our visibility fitting in \S~\ref{subsec:mcmc} are 1.70~$\pm$~0.06 and 13.43~$\pm$~0.13~mJy, respectively, and that at band~7 taken from \citet{vandermarel18a} is 33~$\pm$~0.4~mJy. Note that because our visibility fitting for band~7 data shows large uncertainty (Figure~\ref{figA6}), we used the flux density reported in the literature. The value of $\alpha$ was then calculated as 2.42~$\pm$~0.04, which is lower than the value of $\alpha \sim$2.7 found in other transitional disk systems \citep{pini2014a}. This indicates that larger dust grains are present in the disk or that the disk tends to be partially optically thick at (sub-)millimeter wavelengths.

Figure~\ref{fig1}(c) shows radial cuts of the dust continuum (Figure~\ref{fig1}a) and the $^{12}$CO moment~0 map (Figure~\ref{fig1}b) at a PA of 165.71$\degr$ taken from our modeling results in \S~\ref{subsec:mcmc}. Their brightness profiles are similar at $r\lesssim$50~au with a nearly flat slope at $r\lesssim$20~au. In contrast, the two brightness profiles (Figure~\ref{fig1}c) at $r \gtrsim$50~au deviate, possibly due to the larger size of the CO disk \citep{trap2020a} or the presence of small dust grains, which is discussed in \S~\ref{subsec:mcrt}. 

\begin{figure}[ht!]
\begin{centering}
\includegraphics[clip,width=\linewidth]{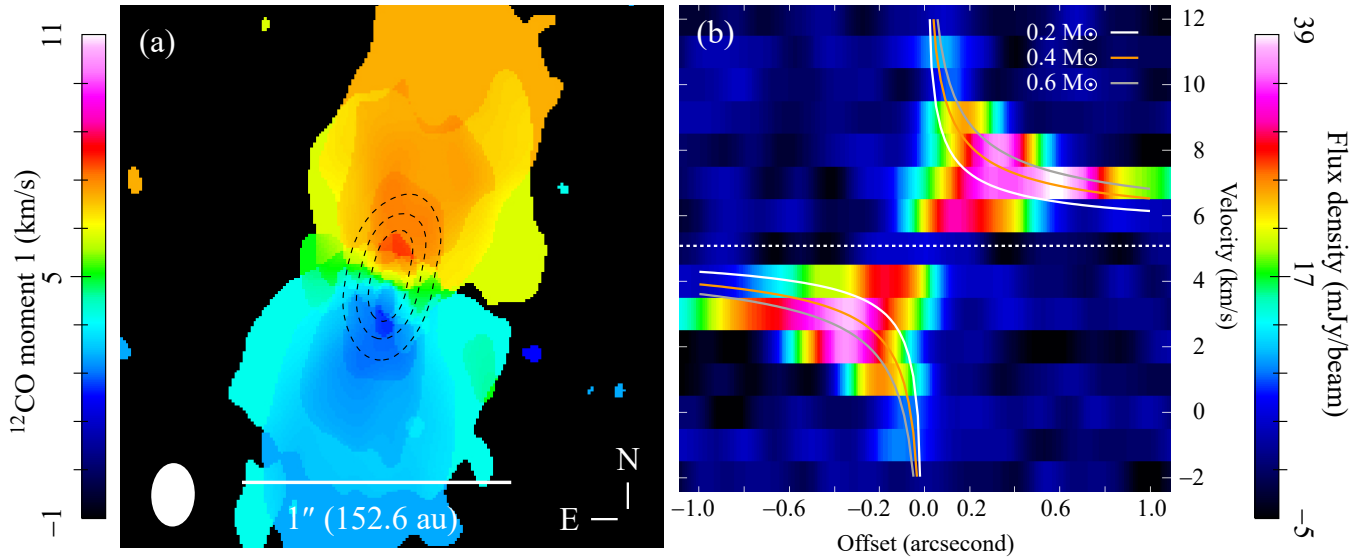}
\end{centering}
\caption{
(a) $^{12}$CO moment~1 map. The contours represent the dust continuum at 20, 40, and 60~$\sigma$. 
(b) PV diagram at a PA of 165.71$\degr$ and a disk inclination of 75.13$\degr$ taken from \S~\ref{subsec:mcmc}. The three lines denote loci of the peak emission of a Keplerian disk around 0.2 to 0.6~$M_{\odot}$ stars. The systemic velocity (white dotted line) is 5.2~km/s \citep{yen2018a}. 
}\label{fig4}
\end{figure}

Figure~\ref{fig4} shows the $^{12}$CO moment~1 map and its position--velocity (PV) diagram at a PA of 165.71$\degr$ and a disk inclination of 75.13$\degr$ taken from \S~\ref{subsec:mcmc}. The center of the CO gas motion in the moment~1 map nearly coincides with the center of the dust continuum emission (Figure~\ref{fig4}a). We plotted the loci of the peak emission of a Keplerian disk around the central star by changing its mass from 0.2 to 0.6~$M_{\odot}$ in the PV diagram and found that the dynamical mass of Sz~84 is $\sim$0.4--0.6~$M_{\sun}$. This value is consistent with the dynamical mass estimated from the $^{13}$CO~$J=3\rightarrow2$ line reported in the literature (0.4~$\pm$~0.1~$M_{\sun}$; \citealp{yen2018a}). Note that these estimates are roughly 2--3 times larger than the spectroscopically determined mass (0.16--0.18~$M_{\sun}$; \citealp{alca2017}). This discrepancy may result from the high disk inclination of 75\degr. $^{12}$CO is generally optically thick and traces the surface layer of the disk, and thus the velocity structure could be largely affected by the high inclination. The disk vertical structure must be taken into account to derive a dynamical mass in the inclined disk system.

\section{Modeling} \label{sec:model}

In this section, we determine the radial structure of the dust disk around Sz~84 by conducting both visibility and SED fitting. Because the high disk inclination of 75\degr \citep{ansd2016a} makes characterizing a cavity with a radius of a few tens of au in an image challenging, modeling is necessary. As mentioned in the introduction (\S~\ref{sec:intro}), (sub-)millimeter and near- to mid-IR observations are generally sensitive to large and small dust grains, respectively, and hence two models, one each for visibility and the SED, were used to determine the radial distributions of large and small dust grains, respectively. Note that \citet{vandermarel18a} derived these models using ALMA cycle~2 band~7 data for Sz~84. Modeling with our new dataset, which has better sensitivity and a better spatial resolution, could better constrain the structures of the dust disk.

\subsection{Visibility fitting} \label{subsec:mcmc}

\begin{figure}[ht!]
\begin{centering}
\includegraphics[clip,width=\linewidth]{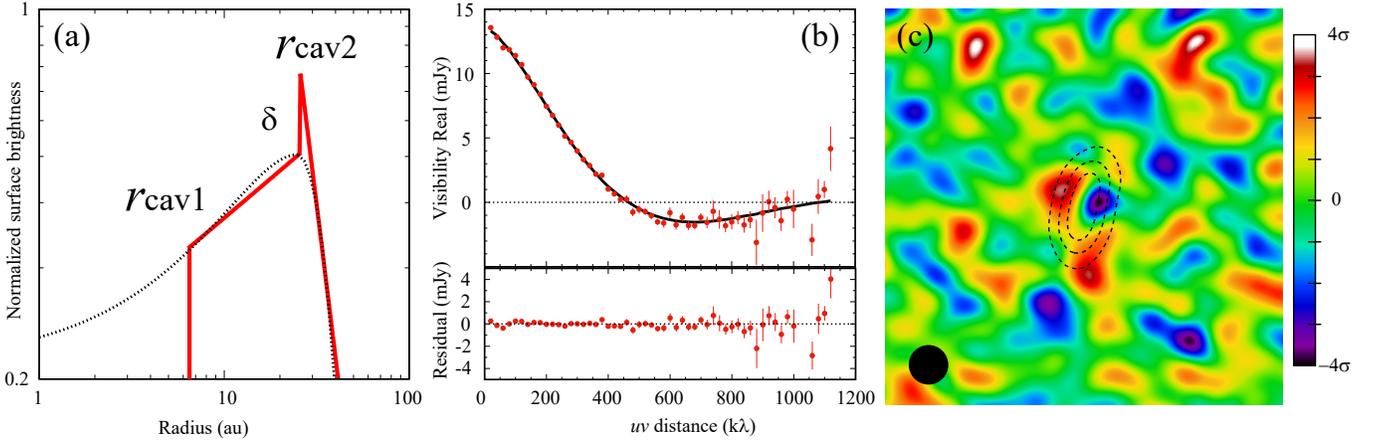}
\end{centering}
\caption{
(a) Surface brightness profile for the best-fit model with notes regarding our visibility fitting (red line). Black dotted line represents an asymmetric Gaussian profile fit by eye to the best-fit profile. (b) Top panel shows real part of the visibilities for the observations (red dots) and the best-fit model (black line); bottom panel shows residual visibilities between observations and the best-fit model. (c) Residual image (2$\arcsec$~$\times$~2$\arcsec$). Black dotted contours represent the dust continuum at SNRs of 20, 40, and 60. 
}\label{fig2}
\end{figure}

The visibilities of Sz~84 show a null point at $\sim$450~k$\lambda$ (Figure~\ref{fig2}b), which suggests cavity (or gap) structures in the protoplanetary disk \citep[c.f.,][]{zhan16}. To confirm the cavity structures, we performed visibility fitting, in which observed visibilities are reproduced with a parametric model of the disk by utilizing all the spatial frequency information.\footnote{We originally attempted to derive the disk structure from the band~3, 6, and 7 data simultaneously to map the spatial distribution of the spectral index $\alpha$. However, we mainly show the results from band~6 in this paper due to the insufficient baseline lengths and lower sensitivities of the band~3 and 7 data (Figures~\ref{figA5} and \ref{figA6}). Data from all three bands were used only to derive the average value of $\alpha$ for the entire disk in \S~\ref{sec:result}.}

We describe the surface brightness distributions of the disk in our model with a simple power-law radial profile:

\begin{eqnarray}
I(r) \propto C_{1} \cdot (r/r_{0})^{-C_{2}}, 
\end{eqnarray}
where $C_{1}$, $C_{2}$, and $r_{0}$ are a scaling factor, the exponent of the power law, and the normalization factor at $r_{\rm cav2}$, respectively. As shown below, this simple profile is sufficient for reproducing the observations. In the radial direction, we adopt the following scaling factors and exponents:

\begin{eqnarray*}
C_{1} &=& \left \{
\begin{array}{llllllll}
0       &{\rm for}&              & & r  &<& r_{\rm cav1}\\
\delta  &{\rm for}& r_{\rm cav1} &<& r  &<& r_{\rm cav2}\\
1       &{\rm for}& r_{\rm cav2} &<& r  &<& 300\ {\rm au},
\end{array}
\right. \\
\end{eqnarray*} 

\begin{eqnarray*}
C_{2} &=& \left \{
\begin{array}{llllllll}
{\rm NA}    &{\rm for}&              & & r  &<& r_{\rm cav1}\\
\gamma_{1}  &{\rm for}& r_{\rm cav1} &<& r  &<& r_{\rm cav2}\\
\gamma_{2}  &{\rm for}& r_{\rm cav2} &<& r  &<& 300\ {\rm au}.
\end{array}
\right. \\
\end{eqnarray*} 

The best-fit surface brightness profile\footnote{Because the best-fit profile can be rather complex, we attempted to fit the best-fit profile with an asymmetric Gaussian profile \citep[e.g.,][]{pini18a} by eye in Figure~\ref{fig2}(a). We found that the asymmetric Gaussian profile extends inside the inner cavity at $r_{\rm cav1}$, which results in significant IR excess, as shown in model~C in \S~\ref{subsec:mcrt}. Thus, we utilized the power-law radial profile in this paper.} with explanations is shown in Figure~\ref{fig2}(a). Because the SED of Sz~84 shows negligible NIR excess, we set no emissions at $r < r_{\rm cav1}$; that is, $C_1=0$. The total flux density ($F_{\rm total}$) was set as a free parameter. The disk inclination ($i$) and PA were also set as free parameters. The phase center was fixed. In total, there are 8 free parameters in our model ($F_{\rm total}$, $r_{\rm cav}$, $r_{\rm gap}$, $\delta$, $\gamma_{1}$, $\gamma_{2}$, $i$, PA)\footnote{The value of $\delta$ was set to zero in the initial fitting. However, we immediately found that the model with $\delta=$0 produces visibilities with an excessively deep gap at the $uv$-distance of $\sim$600k$\lambda$ in Figure~\ref{fig2}(b) and cannot well reproduce the observed visibilities. Therefore, we decided to set $\delta$ as a free parameter.}.

The modeled disk image was converted into complex visibilities with the public Python code \verb#vis_sample# \citep{loomis+17}, in which modeled visibilities are sampled with the same ($u$, $v$) grid points as those in our observations. The modeled visibilities were deprojected\footnote{Visibilities were deprojected in the $uv$-plane using the following equations \citep[e.g.,][]{zhan16}: 
 $u'    =   (u\,{\rm cos\,PA} - v\,{\rm sin\,PA}) \times {\rm cos}\,i,
  v'    =   (u\,{\rm sin\,PA} - v\,{\rm cos\,PA}),$
where $i$ and PA are free parameters in our visibility analyses in \S~\ref{subsec:mcmc}.} with the system PA and $i$ as free parameters, and were calculated as azimuthal mean values ($V_{\rm mean}$) and standard deviations ($\sigma$) within 20~k$\lambda$ bins in the real part. The fitting was performed with the Markov chain Monte Carlo (MCMC) method in the \verb#emcee# package \citep{foreman-mackey+2013}. The log-likelihood function ln$L$ in the MCMC fitting was 
${\rm ln}L = -0.5 \Sigma \{({\rm Re}V_{{\rm mean},j}^{\rm obs} - {\rm Re}V_{{\rm mean},j}^{\rm model})^{2}/\sigma_{j}^{2}\}$, where ${\rm Re}V_{{\rm mean},j}^{\rm obs}$, ${\rm Re}V_{{\rm mean},j}^{\rm model}$, and $\sigma_{j}$ are the observed and modeled visibilities and standard deviations in the real part, respectively. The subscript $j$ represents the $j$-th bin. Our calculations used flat priors with the parameter ranges summarized in Table~\ref{tab:table1}. The burn-in phase (from initial conditions to reasonable sampling) employed 2000 steps, and we ran another 2000 steps for convergence, for a total of 4000 steps with 200 walkers. These fitting procedures were also applied to the band~3 and 7 data with the same parameter ranges except for that of the total flux density. However, we found multiple peaks and a nearly flat posterior probability distribution, especially for the disk inclination and flux density in band~3 and 7 data, respectively, possibly due to shorter baseline lengths and lower sensitivities (Table~\ref{tab:table4} and Figures~\ref{figA5} and \ref{figA6} in the Appendix). Therefore, we decided to use only the flux density from band~3 data (i.e., band~7 data were not used).

The fitting results with errors computed from the 16th and 84th percentiles are shown in Table~\ref{tab:table2}, the radial profile of best-fit surface brightness is shown in Figure~\ref{fig2}(a), the best-fit visibilities with a reduced-$\chi^{2}$ of 1.24 are shown in Figure~\ref{fig2}(b), the best-fit modeled image is shown in Figure~\ref{figA3}, and the probability distributions for the MCMC posteriors are shown in Figure~\ref{figA2}. We found that the cavity at $r=$26~au (i.e., $r_{\rm cav2}$) is shallow with $\delta \gtrsim$0.2 (Figure~\ref{figA2}). We also found that it is difficult to determine the size of the empty cavity (i.e., $r_{\rm cav1}$) due to insufficient baselines in our observations, which have an upper limit of $r_{\rm cav1}\sim$10~au (Figure~\ref{figA2}). A better constraint on this size can be obtained by the SED fitting (see \S~\ref{subsec:mcrt}). The flux inside the 26~au cavity was measured at 4.4~mJy in the best-fit model disk, which is 33~\% of the total flux at band~6. We subtracted the modeled visibilities from the observed ones, and made a CLEANed image (Figure~\ref{fig1}c). 

In \S~\ref{sec:result}, we mentioned the similarity of the radial profiles at $r\lesssim$50~au in the dust continuum and $^{12}$CO moment~0 images. As shown here, the dust continuum emission decreases at $r<$26~au. Hence, the brightness of $^{12}$CO might decrease in the central region of the disk as well. Note that we assumed that the flux attenuation in the central region of the CO gas disk caused by the foreground cloud absorption is negligible due to the high inclination of Sz~84 (75.1$\degr$).

\begin{deluxetable*}{ccccccccc}
\tabletypesize{\footnotesize}
\tablewidth{0pt} 
\tablenum{2}
\tablecaption{Results of MCMC fitting for band~6 data and corresponding parameter ranges \label{tab:table2}}
\tablehead{
\colhead{Flux} & \colhead{$r_{\rm cav1}$} & \colhead{$r_{\rm cav2}$} & \colhead{$\gamma_{1}$} & \colhead{$\gamma_{2}$} & \colhead{$\delta$} & \colhead{$i$} & \colhead{P.A.} & \colhead{Reduced $\chi^{2}$} \\
\colhead{(mJy)} & \colhead{(au)}              & \colhead{(au)}  & \colhead{}   & \colhead{} &\colhead{}& \colhead{($\degr$)} & \colhead{($\degr$)} & \colhead{}
}
\colnumbers
\startdata
13.43$^{+0.13}_{-0.13}$ & 6.45$^{+4.12}_{-4.29}$ &  25.90$^{+2.01}_{-1.85}$ & $-0.43^{+0.85}_{-0.46}$ & 4.07$^{+0.16}_{-0.13}$ & 0.59$^{+0.26}_{-0.25}$ & 75.13$^{+1.74}_{-0.46}$ & 165.71$^{+3.22}_{-2.58}$ & 1.24 \\
\{10.0-20.0\} & \{0.0 .. 21.4\} & \{15.3 .. 45.8\} & \{0.0 .. 10\} & \{-10 .. 10\} & \{0.0 .. 1.0\} & \{50 .. 80\} & \{150 .. 180\} & \\
\hline
\enddata
\tablecomments{
 Values in parentheses are parameter ranges in our MCMC calculations. }
\end{deluxetable*}

\subsection{SED fitting} \label{subsec:mcrt}

\begin{figure}[ht!]
\begin{centering}
\includegraphics[clip,width=\linewidth]{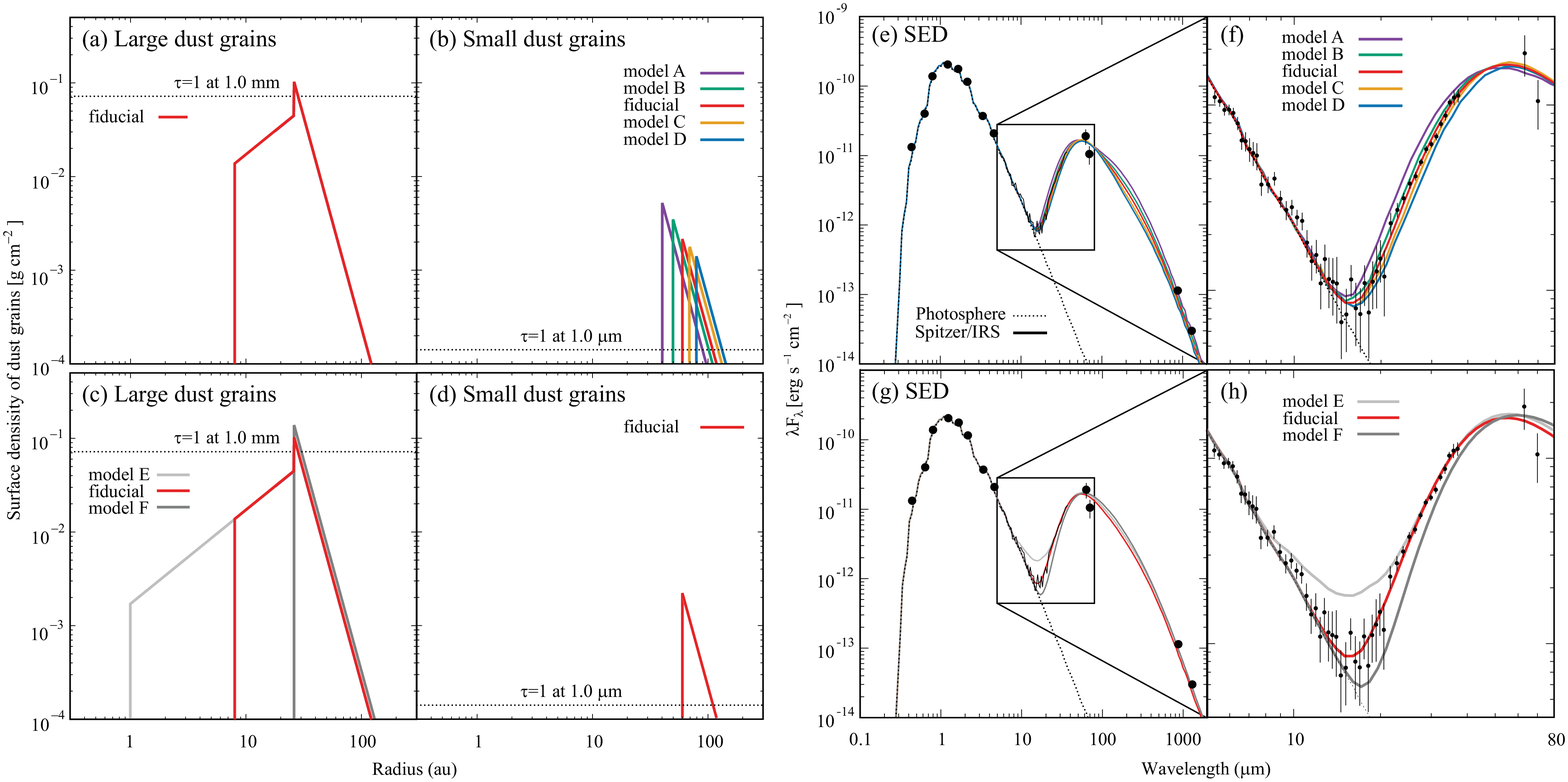}
\end{centering}
\caption{Results of MCRT modeling. Upper and lower panels show models in which the distributions of small and large dust grains are varied, respectively. (a to d): Surface density profiles of large (millimeter size) and small (sub-micron size) dust grains. The surface density profiles of large dust grains in the models shown in the upper panels are same, while the surface density profiles of small dust grains in the models shown in the lower panels are same. Horizontal dotted black lines represent the values of $\tau=1$ in the vertical direction at $\lambda=$1.0~mm and 1~$\mu$m in panels (a and c) and (b and d), respectively. (e to h): SEDs for our modeling. The models are described in Table~\ref{tab:table3}.
}\label{fig3}
\end{figure}

In the previous section, our model showed that the innermost cavity ($r_{\rm cav1}$) is located at $r\lesssim$10~au. Because Sz~84's SED has negligible IR excess at $\lambda \lesssim$10~$\mu$m (Figure~\ref{fig3}e), we need to confirm that our model does not produce significant IR excess. Furthermore, we found a shallow cavity at $r=$26~au in the disk of large dust grains in the previous section. This size is roughly consistent with the $r=20$~au estimated by \citet{vandermarel18a}, who suggested that the 20~au cavity can reproduce the SED of Sz~84. However, this size is smaller than the cavity size derived from the SED analysis by \citet{meri2010a}, namely 55~au in radius. Therefore, we revisited the SED analyses to constrain the radial distributions of both large and small dust grains by running radiative transfer modeling using a Monte Carlo radiative transfer (MCRT) code (\verb#HOCHUNK3D#; \citealp{whit2013}). 

The MCRT code follows a two-layer disk model with small (up to micron size) dust grains in the upper disk atmosphere and large (up to millimeter size) dust grains in the disk midplane \citep[e.g.,][]{dale2006}. The modeled disk structure and dust properties in the MCRT code are described in our previous studies \citep{hashimoto+15}. Briefly, small dust grains are from the standard interstellar-medium dust model (a composite of silicates and graphites with a size distribution of $n(s) \propto s^{-3.5}$ from $s_{\rm min} = 0.0025$~$\mu$m to $s_{\rm max} = 0.2$~$\mu$m) in \citet{kim1994} and large dust grains (a composite of carbons and silicates with a size distribution of $n(s) \propto s^{-3.5}$ from $s_{\rm min} = 0.01$~$\mu$m to $s_{\rm max} = 1000$~$\mu$m) are from Model~2 in \citet{wood2002}. The radial surface density was assumed to have a simple power-law radial profile similar to the equation in \S~\ref{subsec:mcmc}:  

\begin{eqnarray*}
\Sigma(r) &=& C_{3} \cdot \Sigma_{0} (r/r_{0})^{-C_{4}}, \\
C_{3} &=& \left \{
\begin{array}{llllllll}
0       &{\rm for}&              & & r  &<& r_{\rm cav1}\\
\delta  &{\rm for}& r_{\rm cav1} &<& r  &<& r_{\rm cav2}\\
1       &{\rm for}& r_{\rm cav2} &<& r  &<& 300\ {\rm au}
\end{array}
\right. 
\end{eqnarray*} 

\begin{eqnarray*}
C_{4} &=& \left \{
\begin{array}{llllllll}
{\rm NA}    &{\rm for}&              & & r  &<& r_{\rm cav1}\\
q_{1}  &{\rm for}& r_{\rm cav1} &<& r  &<& r_{\rm cav2}\\
q_{2}  &{\rm for}& r_{\rm cav2} &<& r  &<& 300\ {\rm au}.
\end{array}
\right. \\
\end{eqnarray*} 
where $r_{0}$ is the normalization factor at $r_{\rm cav2}$, $\Sigma_{0}$ is the normalized surface density determined from the total (gas~$+$~dust) disk mass ($M_{\rm disk}$) assuming a gas-to-dust mass ratio of 100, $C_{3}$ is the scaling factor for the surface density (which is set to the same value as $C_{1}$ in \S~\ref{subsec:mcmc} for large dust grains), and $C_{4}$ is the radial gradient parameter. The scaling factor for small dust grains ($\delta^{\rm small}$) was assumed to be zero at $r<r^{\rm small}_{\rm rcav2}$. The radial gradient parameters $q_{1}$ and $q_{2}$ (which are the same for disks of large and small dust grains) were fixed to reproduce the surface brightness derived in \S~\ref{subsec:mcmc}, as explained below. The value of $r^{\rm large}_{\rm rcav2} =$26~au was taken from the result in \S~\ref{subsec:mcmc}. $r^{\rm large}_{\rm rcav1}$ and $r^{\rm small}_{\rm rcav2}$ were free parameters. We set $M_{\rm disk}$ to reproduce a flux of 13.43~mJy at 1.3~mm. The scale heights ($h$) of large and small grains were assumed to vary as $h \propto r^{p}$. For simplicity, we assumed $p=$ 1.25 with a typical midplane temperature profile of $T \propto r^{-0.5}$. We fixed the scale heights of 1~au at $r=$ 100~au for the disk of large dust grains ($h^{\rm large}_{r=100{\rm au}}$). Those of small dust grains ($h^{\rm small}_{r=100{\rm au}}$) were varied to reproduce the flux at $\lambda \sim$60--70~$\mu$m. The mass fraction ($f$) of large dust grains in the total dust mass was set to 0.9. The disk inclination was set to 75$\degr$, as derived in our visibility analyses (\S~\ref{subsec:mcmc}). The \verb#HOCHUNK3D# code calculates the accretion luminosity from the star based on the mass accretion rate. Half of the flux was emitted as X-rays (which heat the disk) and half as stellar flux at a higher temperature. We set a mass accretion rate of $\dot{M} =$~1.3~$\times $10$^{-9}$~$M_{\odot}$/yr \citep{mana14}. In the code, we varied four parameters, namely $r^{\rm large}_{\rm rcav1}$, $r^{\rm small}_{\rm rcav2}$, $h^{\rm small}_{r=100{\rm au}}$, and $M_{\rm disk}$, as shown in Table~\ref{tab:table3}, along with the $\chi^{2}$ calculated at $\lambda=$1--100~$\mu$m.

Figure~\ref{fig3} shows the surface densities and the SEDs for each model. Photometry and spectroscopy data used here are summarized in Table~\ref{tab:phot}. Our fitting procedure included the following four steps.

\begin{enumerate}
\item We first tentatively set $r^{\rm large}_{\rm rcav1} =$10~au, $q_{1}=-$0.5, $q_{2}=$4.0 (which are derived in \S~\ref{subsec:mcmc}) and $r^{\rm small}_{\rm rcav2} =$55~au (which is taken from \citealp{meri2010a}) with varying values of $h^{\rm small}_{r=100{\rm au}}$ and $M_{\rm disk}$, and found 
that the values of $h^{\rm small}_{r=100{\rm au}}=$5.0~au and $M_{\rm disk}=$2.5~$M_{\rm Jup}$ well reproduce far-IR flux at $\lambda \sim$60--70~$\mu$m and millimeter flux at 1.3~mm in the SED (Figure~\ref{fig3}e to g). 

\item We iteratively varied the radial gradient parameters $q_{1}$ and $q_{2}$ to reproduce the radial surface brightness of the dust continuum at band~6 derived in \S~\ref{subsec:mcmc}, and found that the values of $q_{1}=-$1.0 and $q_{2}=$4.5 (fiducial model) well reproduce the surface brightness, as shown in Figures~\ref{figA4}(a) and (b). The bumps at $r\sim$10 and $\sim$60~au in the surface brightness of the MCRT modeling are attributed to those of the midplane temperature in Figure~\ref{figA4}(c) due to irradiated cavity walls. We fixed these two values in the following procedures.

\item To constrain the cavity size of small dust grains, $r^{\rm small}_{\rm rcav2}$ was set to 40 (model~A), 50 (model~B), 60 (fiducial model), 70 (model~C), and 80~au (model~D), as shown in upper panels in Figure~\ref{fig3}, with iterative adjustment of $h^{\rm small}_{r=100{\rm au}}$. The SED can be well reproduced by a cavity in the small dust with a size of $60\pm10$ au.

\item Finally, we set $r^{\rm large}_{\rm rcav1}$ at 1 (model~E), 8 (fiducial model), and 26~au (model~F), as shown in lower panels in Figure~\ref{fig3}, to constrain the size of the inner cavity in the disk of large dust grains. The 8~au cavity, which is consistent with the results of MCMC modeling in \S~\ref{subsec:mcmc} (i.e., $r^{\rm large}_{\rm rcav1} \lesssim$10~au), is suitable for the fiducial model. 
\end{enumerate}

In summary, our modeling suggests that the small dust grains are located at $r\gtrsim$60~au and the large dust grains are distributed at 10~au~$\lesssim r \lesssim$~60~au. There is a dust-free region at $r\lesssim$10~au. 

Because we used two independent disk models in the visibility fitting (\S~\ref{subsec:mcmc}) and the SED fitting (\S~\ref{subsec:mcrt}), we compare the surface brightness of the two models at band~6 in Figure~\ref{figA4} to check their consistency. The maximum residual between the two surface brightness profiles convolved with the ALMA observation beam of 0\farcs2 (30~au) in Figure~\ref{figA4}(b) is $\sim$119~$\mu$Jy/beam (1.9~$\sigma$). This residual is insignificant and thus the two models are mutually consistent.

Our picture of the disk with small dust grains residing only at large radii is supported by gas observations.  As shown in \S~\ref{sec:result}, the radial profile of the CO moment~0 deviates from that of dust at $r\gtrsim 50$~au. Small dust grains are the main source of heating as they are present in the upper layer of the disk and are thus directly irradiated by the stellar light and effectively absorb it. Therefore, the outer disk is heated more effectively than the inner disk, which contains only a small amount of small dust grains. Because small dust grains do not emit effectively at (sub-)millimeter wavelengths and the amount of large dust grains rapidly decreases at outer radii ($\propto r^{-4.5}$), these two facts do not greatly affect dust continuum emission.  However, the $^{12}$CO gas emission mainly traces the temperature of the disk and therefore its moment~0 emission profile should be affected by the presence of small dust grains.

\begin{deluxetable*}{cccccccccccccccc}
\tablewidth{0pt} 
\tablenum{3}
\tablecaption{Parameters in our MCRT modeling \label{tab:table3}}
\tablehead{
\colhead{Model} & \colhead{$r_{\rm cav1}^{\rm large}$} & \colhead{$r_{\rm cav1}^{\rm small}$} & \colhead{$r_{\rm cav2}^{\rm large}$}& \colhead{$r_{\rm cav2}^{\rm small}$}  & \colhead{$\delta^{\rm large}$} & \colhead{$\delta^{\rm small}$} & \colhead{$p$} & \colhead{$q_{1}$} & \colhead{$q_{2}$} & \colhead{$h^{\rm large}_{r=100au}$} & \colhead{$h^{\rm small}_{r=100au}$} & \colhead{$M_{\rm disk}$} & \colhead{$\dot{M}$} & \colhead{$i$} & \colhead{$\chi^{2}$}\\
\colhead{} & \colhead{(au)} & \colhead{(au)} & \colhead{(au)} & \colhead{(au)}  & \colhead{}    & \colhead{} & \colhead{} & \colhead{}   & \colhead{} & \colhead{(au)}     & \colhead{(au)} & \colhead{($M_{\rm Jup}$)} & \colhead{($M_{\sun}$~yr$^{-1}$)} & \colhead{(\degr)} & \colhead{(\degr)}
}
\colnumbers
\startdata
fiducial &  8.0  & --- & 26.0  &  60.0 & 0.59 & 0    & 1.25 & $-$1.0 & 4.5 & 1.0 & 5.0 & 2.5 & 1.3~$\times$~10$^{-9}$ & 75 & 151.0\\
A        &  8.0  & --- & 26.0  &  40.0 & 0.59 & 0    & 1.25 & $-$1.0 & 4.5 & 1.0 & 5.8 & 2.5 & 1.3~$\times$~10$^{-9}$ & 75 & 662.4\\
B        &  8.0  & --- & 26.0  &  50.0 & 0.59 & 0    & 1.25 & $-$1.0 & 4.5 & 1.0 & 5.2 & 2.5 & 1.3~$\times$~10$^{-9}$ & 75 & 200.0\\
C        &  8.0  & --- & 26.0  &  70.0 & 0.59 & 0    & 1.25 & $-$1.0 & 4.5 & 1.0 & 4.8 & 2.5 & 1.3~$\times$~10$^{-9}$ & 75 & 180.7\\
D        &  8.0  & --- & 26.0  &  80.0 & 0.59 & 0    & 1.25 & $-$1.0 & 4.5 & 1.0 & 4.5 & 2.5 & 1.3~$\times$~10$^{-9}$ & 75 & 299.6\\
E        &  1.0  & --- & 26.0  &  60.0 & 0.59 & 0    & 1.25 & $-$1.0 & 4.5 & 1.0 & 5.5 & 2.5 & 1.3~$\times$~10$^{-9}$ & 75 & 312.4\\
F        & 26.0  & --- & 26.0  &  60.0 & 0.59 & 0    & 1.25 & $-$1.0 & 4.5 & 1.0 & 5.0 & 2.5 & 1.3~$\times$~10$^{-9}$ & 75 & 478.2\\
\enddata
\end{deluxetable*}

\subsection{Upper limit on amount of small grains inside cavity} \label{subsec:limit}

Following \citet{beichman05}, we converted the upper limit on the measured IR ($\lambda\lesssim$10~\micron) flux at a certain wavelength $\lambda$ into an upper limit on the amount of small dust grains inside the 60~au cavity, $M_{\rm cav}^{\rm small}$. At NIR wavelengths, the emission from large dust grains at $r>$26~au is negligible, as shown in our SED fitting (\S~\ref{subsec:mcrt}). The value of $M_{\rm cav}^{\rm small}$ is expected to be very small because the NIR flux is consistent with stellar photospherical emission. We assumed that the excess emission at NIR wavelengths is optically thin (i.e., a transitional disk with a cavity, \citealt{espaillat10}). The surface density of small dust grains inside the cavity, $\Sigma_{\rm cav}^{\rm small}$, was assumed to vary with the radius as 
\begin{eqnarray*}
\Sigma_{\rm cav}^{\rm small}(r)=\Sigma_{\rm cav}^{\rm small}(r_0) (r/r_0)^\alpha \ 
\mbox{for } r^{\rm min} < r < r^{\rm max}
\end{eqnarray*} 
 where $r_0=$1~au and $\Sigma_{\rm cav}^{\rm small}(r_0)$ is the surface density at 1~au. The total dust mass is thus 
\begin{eqnarray*}
 M_{\rm cav}^{\rm small} = \int_{r^{\rm min}}^{r^{\rm max}} \Sigma_{\rm cav}^{\rm small}(r) 2\pi r dr. 
\end{eqnarray*} 
We calculated the equilibrium temperature of optically thin dust at each radius inside the cavity, $T(r)$, using the \verb#HOCHUNK3D# code. The total NIR flux from the cavity dust at a given wavelength is
\begin{eqnarray*}
 I(\lambda)=\int_{r_{\rm min}}^{r_{\rm max}} B(T(r),\lambda) \kappa_{\rm small}(\lambda) \Sigma_{\rm cav}^{\rm small}(r) 2\pi r dr, 
\end{eqnarray*} 
where $B(T(r),\lambda)$ is the Planck function at $T(r)$ and $\lambda$, and $\kappa_{\rm small}(\lambda)$ is the absorption opacity of small dust grains. Note that because our purpose is to derive the total mass of small dust grains from the upper limit of the observed flux, disk inclination is not needed to take into account in the calculation.

At $\lambda=$10~\micron, the upper limit on the IR excess of the system $\lambda F(\lambda)$ is $2.5\times10^{-13}$~ergs~s$^{-1}$~cm$^{-2}$. Assuming that $r^{\rm min}$ is the dust sublimation radius at 0.03~at, $r^{\rm max}$ is 60~au, and $\alpha=-1$ (self-similar viscous disk solution with viscosity $\propto$ radius, \citealt{lyndenbell74}), the upper limit on $M_{\rm cav}^{\rm small}$ is $4\times10^{-7}M_\oplus$ ($\Sigma^{\rm small}_{r=26 {\rm au}}=1.2\times10^{-9}$ g~cm$^{-2}$, equivalent to $\delta^{\rm small}=4\times10^{-7}$). We note that the value of $M_{\rm cav}^{\rm small}$ is only weakly affected by the chosen wavelength. For example, using $\lambda=$3.35~$\micron$, the associated upper limit on the disk flux $\lambda F(\lambda)<7.9\times10^{-13}$~ergs~s$^{-1}$~cm$^{-2}$ yields $M_{\rm cav}^{\rm small} = 1\times10^{-6}M_\oplus$. Decreasing $r_{\rm max}$ reduces $M_{\rm cav}^{\rm small}$ (e.g., setting $r_{\rm max}=$5~au results in $M_{\rm cav}^{\rm small}=4\times10^{-8}M_\oplus$) and increasing $r_{\rm min}$ has the opposite effect (e.g., setting $r_{\rm min}=$1~au results in $M_{\rm cav}^{\rm small}=8\times10^{-6}M_\oplus$). Flattening or reversing the dependence of $\Sigma_{\rm cav}^{\rm small}(r)$ on radius yields a higher value of $M_{\rm cav}^{\rm small}$ (e.g., setting $\alpha=1$ results in $M_{\rm cav}^{\rm small}=1\times10^{-3}M_\oplus$).

We conclude that the amount of small dust inside the millimeter emission cavity is tiny, less than a lunar mass under reasonable assumptions. The mass of large dust grains at $r<$60~au is $\sim$7~$M_{\earth}$, and thus the ratio of the mass of small dust grains to that of large dust grains is smaller than $\sim$10$^{-4}$.

\section{Discussion} \label{sec:discuss}

\subsection{Comparison with other systems} \label{subsec:comparison}

\begin{figure}[ht!]
\begin{centering}
\includegraphics[clip,width=\linewidth]{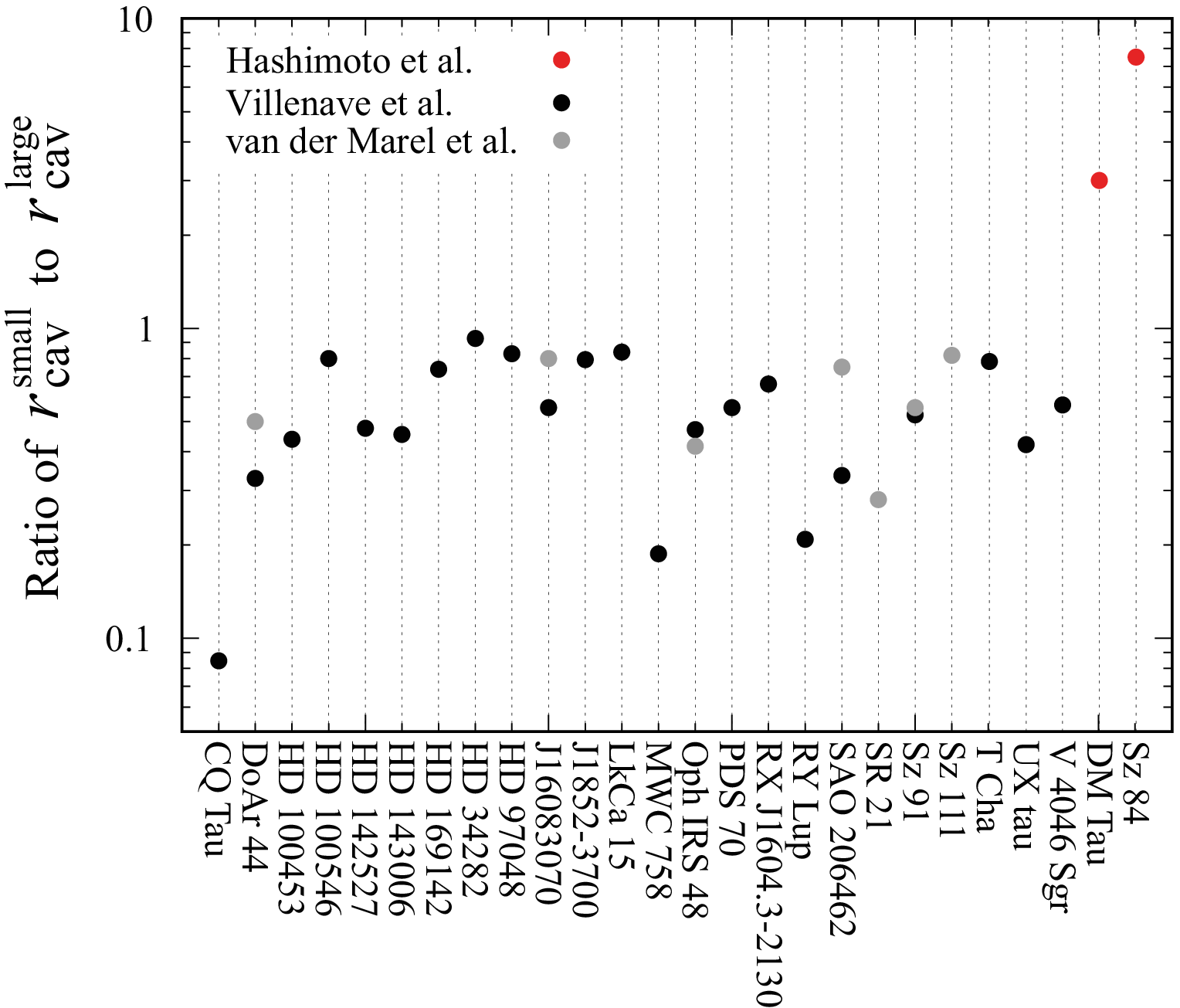}
\end{centering}
\caption{
Cavity ratio of $r_{\rm cav}^{\rm small}$ (small dust grains) to $r_{\rm cav}^{\rm large}$ (large dust grains). Black dots denote the cavity ratios calculated using the NIR and (sub-)millimeter dust continuum observations taken from \citep{vill2019a}, and gray dots are those calculated using the CO gas and dust continuum observations taken from \citep{vandermarel+16,vandermarel18a}. We assumed that the cavity sizes of the gas and small dust grains are identical. In the samples from \citet{vandermarel18a}, we use three objects with constrained cavity sizes of CO gas. Red dots represent our two objects (Sz~84 from this work and DM~Tau from Hashimoto et~al. submitted) with $r_{\rm cav}^{\rm small} > r_{\rm cav}^{\rm large}$ mentioned in \S~\ref{subsec:comparison}.
}\label{fig5}
\end{figure}

The sizes of the cavities of large and small dust grains have been discussed by combining NIR direct imaging and radio interferometry \citep[e.g.,][]{dong12cavity,vill2019a}. Additionally, assuming that small dust grains are well coupled to the gas, measured CO gas distributions might serve as a proxy of the spatial distribution of small dust grains. CO gas observations with sufficient spatial resolution and sensitivity have been conducted with ALMA \citep[e.g.,][]{vandermarel+15b,vandermarel+16,vandermarel18a}. The general trend of the cavity sizes in the disk of large dust grains vs. small ones (or CO gas) indicates that the cavity size in the disk of large dust grains is largest. The ratios of $r_{\rm cav}^{\rm small}$ to $r_{\rm cav}^{\rm large}$ and those of $r_{\rm cav}^{\rm CO}$ to $r_{\rm cav}^{\rm large}$ from the literature are compiled in Figure~\ref{fig5}. These results can be interpreted as outcomes of planet--disk interactions \citep[e.g.,][]{zhu12,dujuanovelar2013,facc2018b}, as described in \S~\ref{sec:intro}. 

For Sz~84, the cavity sizes show the opposite trend (Figure~\ref{fig5}), i.e., the disk composed of small dust grains has a larger cavity, as shown in \S~\ref{sec:model}. As far as we know, such structures can only be found in DM~Tau ($r_{\rm cav}^{\rm small} \sim$3~au and $r_{\rm cav}^{\rm large} \sim$1~au; Hashimoto et~al. submitted). Systems with a larger cavity in the disk of small dust grains may be relatively rare. \citet{vill2019a} and \citet{vandermarel+16,vandermarel18a} compared the cavity sizes ($r_{\rm cav}^{\rm small}$ vs. $r_{\rm cav}^{\rm large}$ or $r_{\rm cav}^{\rm CO}$ vs. $r_{\rm cav}^{\rm large}$) for 24 objects and found that all the disks have larger cavities of large dust grains (Figure~\ref{fig5}). There are currently two objects (Sz~84 and DM~Tau) that have a larger cavity in the disk of small dust grains, accounting for 7.7~\% of objects observed with NIR direct imaging and ALMA. However, current NIR observations are biased toward brighter objects ($R$~band $\lesssim$12~mag) due to the limitations of the adaptive optics system. Future observations with extreme adaptive optics systems such as VLT/SPHERE and Subaru/SCExAO would increase the number of samples (especially faint objects) and thus further constrain the statistics.

\subsection{Origin of cavities}

Although systems with large dust grains inside the cavity in the disk of small dust grains may be rare, Sz~84 data have important implications for understanding planet formation. Here, we discuss the possible origins of such cavity structures. Some formation mechanisms of the cavity have been proposed, including photoevaporation \citep[e.g.,][]{clar2001a,alex2014}, planet--disk interactions \citep[e.g.,][]{kley12}, grain growth \citep[e.g.,][]{dull2005a}, radiation pressure \citep[e.g.,][]{krum2019a}, and a combination of these mechanisms (e.g., planet $+$ grain growth; \citealp{zhu12}).

\emph{Photoevaporation ---} The basic idea of photoevaporation is that high-energy radiation (UV/X-ray) from the central star heats the surface of the disk, and the hot gas  ($\sim$10$^{3}$--10$^{4}$~K) escapes from the disk as a photoevaporative wind. Because small dust grains are well coupled to the gas, they flow with the gas, possibly resulting in similar sizes of the cavities in the disks of small dust grains and gas. However, $^{12}$CO emissions can be clearly seen at $r<$60~au (Figure~\ref{fig1}). Furthermore Sz~84 has a moderate mass accretion rate \citep{mana14}. Thus, photoevaporation does not seem to be a viable mechanism for creating the cavity in the disk around Sz~84 \citep[e.g.,][]{erco2017a}. 

\emph{Planet--disk interactions ---} This mechanism is commonly considered as a potential origin of transitional disks \citep[e.g.,][]{espaillat+14}. However, the robust detection of young planets that possibly induce cavity/gap structures is currently very limited \citep[e.g., PDS~70b \& c;][]{kepp18a,haff19a}, and no planets have been detected around Sz~84. Additionally, as mentioned above (\S~\ref{subsec:comparison}), the spatial distributions of small and large dust grains expected for planet--disk interactions are opposite to those around Sz~84. Furthermore, as discussed in \citet{zhu12}, because gas ($+$ small dust grains) can flow across the planet-induced gap, NIR excess in the SED emitted by hot ($\gtrsim$100--1000~K) dust grains around the central star cannot be sufficiently suppressed by only planet--disk interactions. This is also opposite to Sz~84's SED with negligible NIR excess. Therefore, planet--disk interactions alone cannot account for the observational results for Sz~84.  

\emph{Grain growth ---} To account for a transitional disk system with negligible NIR excess in the SED and a moderate mass accretion, only small dust grains in the vicinity of the central star need to be removed. For this purpose, \citet{zhu12} introduced grain growth in addition to planet--disk interactions. Dust grains grow faster in the inner region of the disk because of a faster dynamical time scale \citep[e.g.,][]{brauer+2008}, and thus small dust grains are dispersed from the inside out. However, large dust grains are expected to fragment into small dust grains depending on disk turbulence \citep[e.g.,][]{weid1984a} and/or their stickiness \citep[e.g.,][]{wada2009}, and thus weak dust fragmentation is required to efficiently remove small dust grains. Numerical simulations of grain growth with no/low fragmentation show a deficit of flux at NIR to MIR wavelengths in the SED \citep[e.g.,][]{dull2005a,birn2012a}. Moreover, as grain growth does not affect gas distribution, the mass accretion would be unchanged. Hence, grain growth with less efficient fragmentation would explain both the spatial distributions of dust grains at $r\gtrsim$10~au and the moderate mass accretion rate of Sz~84. We note that our modeling in \S~\ref{sec:model} implies that the inner disk at $r\lesssim$10~au is optically thin at $\lambda \lesssim$10~$\mu$m for both small and large dust grains, which could be interpreted as meaning that dust grains are large enough to radially drift to the central star \citep[e.g.,][]{brauer+2008} or have already grown into planetesimals \citep[e.g.,][]{okuz2012,kata2013a}. 

As mentioned in \S~\ref{subsec:comparison}, Sz~84 could be a very rare system because most T~Tauri stars have IR excesses in their SEDs \citep[e.g.,][]{espaillat+14}. These facts suggest that in most T~Tauri star systems, small dust grains need to be replenished after being removed by grain growth, for instance by dust fragmentation through high-speed collisions. Under reasonable assumptions, the balance of the rates of grain growth and fragmentation determines whether T~Tauri star systems have IR excess. For Sz~84, grain growth could be very efficient and/or fragmentation could be very inefficient, potentially due to weak turbulence in the disk or stickier dust grains. 

Furthermore, according to numerical simulations in \citet{dull2005a}, the time scale of grain growth with less effective fragmentation required to create the dip at NIR to MIR wavelengths in the SED is expected to be $\lesssim$0.1~Myr, and the cavity size in the disk of small dust grains reaches $r\sim$100~au within 1~Myr (see model F2 in Figure~6). This picture is roughly consistent with Sz~84 ($r_{\rm cav}^{\rm small} =$60~au in \S~\ref{subsec:mcrt}; an age of 1~Myr, \citealp{alca2014}). In contrast, DM~Tau, which also has larger cavities of small dust grains, as mentioned in the previous subsection (\S~\ref{subsec:comparison}), has a cavity size on the order of 1~au even though the cavity's age is $\sim$1~Myr \citep[e.g.,][]{andr2013}. Because the time scale of the cavity formation on the order of $r=$1~au via grain growth is predicted to be $\sim$0.01~Myr \citep{dull2005a}, slow grain growth might occur in DM~Tau's disk. These results suggest that Sz~84 has more efficient grain growth and/or more insufficient fragmentation than those of DM~Tau because the time scale of the cavity formation might also depend on the balance of grain growth and fragmentation. Therefore, Sz~84 could serve as an excellent testbed for investigating grain growth with less efficient fragmentation because dust fragmentation is one of the big problems in planet formation. One way to test the scenario of grain growth is to investigate the spectral index as a function of radius. Future multiple wavelength observations with higher spatial resolution and better sensitivity will guide our understanding of the cavity formation mechanism around Sz~84.

\emph{Radiation pressure ---} Another possible mechanism for the selective removal of small dust grains is radiation pressure \citep[e.g.,][]{krum2019a,owen2019}. Stellar radiation radially and rapidly pushes away small dust grains, whereas large dust grains persist for a longer time, when the radiation pressure force on dust grains is stronger than the drag force of the gas. To suppress this drag force, \citet{krum2019a} reduced the gas density to less than 1~\% of the minimum mass solar nebula \citep[e.g.,][]{weid77,haya1981}. Additionally, because the dust disk needs optically thin conditions to avoid shielding stellar radiation, a dust-to-mass ratio of less than 10$^{-4}$ is preferable. Consequently, radiation pressure could efficiently remove small dust grains on gas-poor systems, i.e., debris disks. Conversely, the protoplanetary disk of Sz~84 is likely to be a gas-rich system, as detected in $^{12}$CO (Figure~\ref{fig1}), and therefore, radiation pressure can be safely ruled out as the dominant mechanism for cavity formation around Sz~84.

\section{Conclusion} \label{sec:conclusion}

We observed the transitional disk around the T~Tauri star Sz~84 at band~6 ($\lambda \sim$1.3~mm) with a spatial resolution of 0$\farcs$2 (30~au) with ALMA. We clearly detected a dust continuum and $^{12}$CO~$J=2\rightarrow1$ line emissions. Although the SED of Sz~84 exhibits obvious deficits of IR excess at $\lambda \lesssim$10~$\mu$m, indicating a cavity structure with a size of $\sim$100~au in diameter, the dust continuum image does not show any cavity structures. In contrast, the observed visibilities of the dust continuum clearly show a null point at the $uv$-distance of $\sim$450~k$\lambda$, suggesting a cavity structure in dust continuum. These observational results motivated us to conduct analyses of the visibilities and the SED to explore the structures of the dust disk. Our main findings are as follows.

\begin{enumerate}
\item The spectral index ($\alpha$) at bands~3, 6, and 7 (0.9 to 3~mm) is 2.42~$\pm$~0.04, which is lower than the value of $\alpha \sim$2.7 found in other transitional disk systems \citep{pini2014a}.

\item The analyses of visibilities and the SED show that the cavity in the disk consisting of small dust grains is located at $r\gtrsim$60~au, and that large dust grains are present inside the cavity in the disk of small dust grains at $r\sim$60~au down to $\sim$10~au. Gas is also present at $r<$60~au in the $^{12}$CO moment~0 map.

\item A transitional disk in which the size of the cavity in the disk of small dust grains is larger than that in the disk of large dust grains may be rare (to our knowledge, only Sz~84 and DM~Tau), accounting for 7.7~\% of objects observed with NIR direct imaging and ALMA \citep{vill2019a,vandermarel+16,vandermarel18a}. Note that we assumed that the cavity sizes in disks of gas and small dust grains are identical in the modeling of CO gas taken from \citet{vandermarel+16,vandermarel18a}.  

\item To account for the observational results for Sz~84 (spatial distributions of dust grains and gas in the disk, SED with negligible IR excess, and moderate mass accretion), grain growth and less efficient fragmentation (i.e., not photoevaporation, planet--disk interactions, or radiation pressure) are the likely mechanisms for cavity formation around Sz~84. Grain growth is thought to be an important first step in planet formation, and dust fragmentation prevents dust grains from growing into larger bodies. Therefore, Sz~84 is a good testbed for investigating grain growth with inefficient fragmentation of dust grains.   

\end{enumerate}

\acknowledgments
We thanks an anonymous referee for a helpful review of the manuscript.
This paper makes use of the following ALMA data: ADS/JAO.ALMA\#2016.1.00571.S, ADS/JAO.ALMA\#2015.1.01301.S, and ADS/JAO.ALMA\#2013.1.00220.S. ALMA is a partnership among ESO (representing its member states), NSF (USA), and NINS (Japan), together with NRC (Canada), NSC and ASIAA (Taiwan), and KASI (Republic of Korea), in cooperation with the Republic of Chile. The Joint ALMA Observatory is operated by ESO, AUI/NRAO, and NAOJ.
This work is based in part on archival data obtained with the Spitzer Space Telescope, which was operated by the Jet Propulsion Laboratory, California Institute of Technology under a contract with NASA. Support for this work was provided by an award issued by JPL/Caltech.
This work was supported by JSPS KAKENHI Grant Numbers 19H00703, 19H05089, 19K03932, 18H05442, 15H02063, and 22000005. Y.H. is supported by the Jet Propulsion Laboratory, California Institute of Technology,
under a contract with NASA.

{\it Software}: \verb#vis_sample# \citep{loomis+17}, 
          \verb#HOCHUNK3D# \citep{whit2013}, 
          \verb#CASA# \citep{mcmu07}, 
          \verb#emcee# \citep{foreman-mackey+2013}

\appendix

Here, we provide additional supporting tables and figures. 
ALMA observations of archive data from bands~3 and 7 used in this paper are summarized in Table~\ref{tab:band3-7}.

\begin{deluxetable*}{lll}
\tabletypesize{\footnotesize}
\tablewidth{0pt} 
\tablenum{4}
\tablecaption{ALMA observations of archive data\label{tab:band3-7}}
\tablehead{
\colhead{} & \colhead{Band 3} & \colhead{Band 7}   
}
\startdata
Observing date (UT)         & 2016.Oct.06                    & 2015.Jun.14        \\
Project code                & 2016.1.00571.S                 & 2013.1.00220.S \\
Time on source (min)        & 6.0                            & 2.2 \\
Number of antennas          & 42                             & 41 \\
Baseline lengths            & 18.6~m to 3.1~km               & 21.4~m to 0.78~km\\
Baseband Freqs. (GHz)       & 90.6, 92.5, 102.6, 104.5       & 328.3, 329.3, 330.6, 340.0, 341.8 \\
Channel width   (MHz)       & 0.98, 0.98, 0.98, 0.98         & 15.6,  0.122, 0.122, 0.244, 0.977\\
Continuum band width (GHz)  & 7.5                            &  4.8 \\
Bandpass calibrator         & J1517$-$2422                   &  J1427$-$4206 \\
Flux calibrator             & J1427$-$4206                   &  Titan \\
Phase calibrator            & J1610$-$3958                   &  J1610$-$3958 \\
New phase center with GAIA  & 15h58m2.50256s, -37d36m3.1157s & 15h58m2.50402s, -37d36m3.0857s \\
\enddata
\tablecomments{
Because bands~3 and 6 data are mainly used for visibility analyses in \S~\ref{subsec:mcmc}, we do not describe the imaging parameters.}
\end{deluxetable*}

\begin{figure}[ht!]
\begin{centering}
\includegraphics[clip,width=\linewidth]{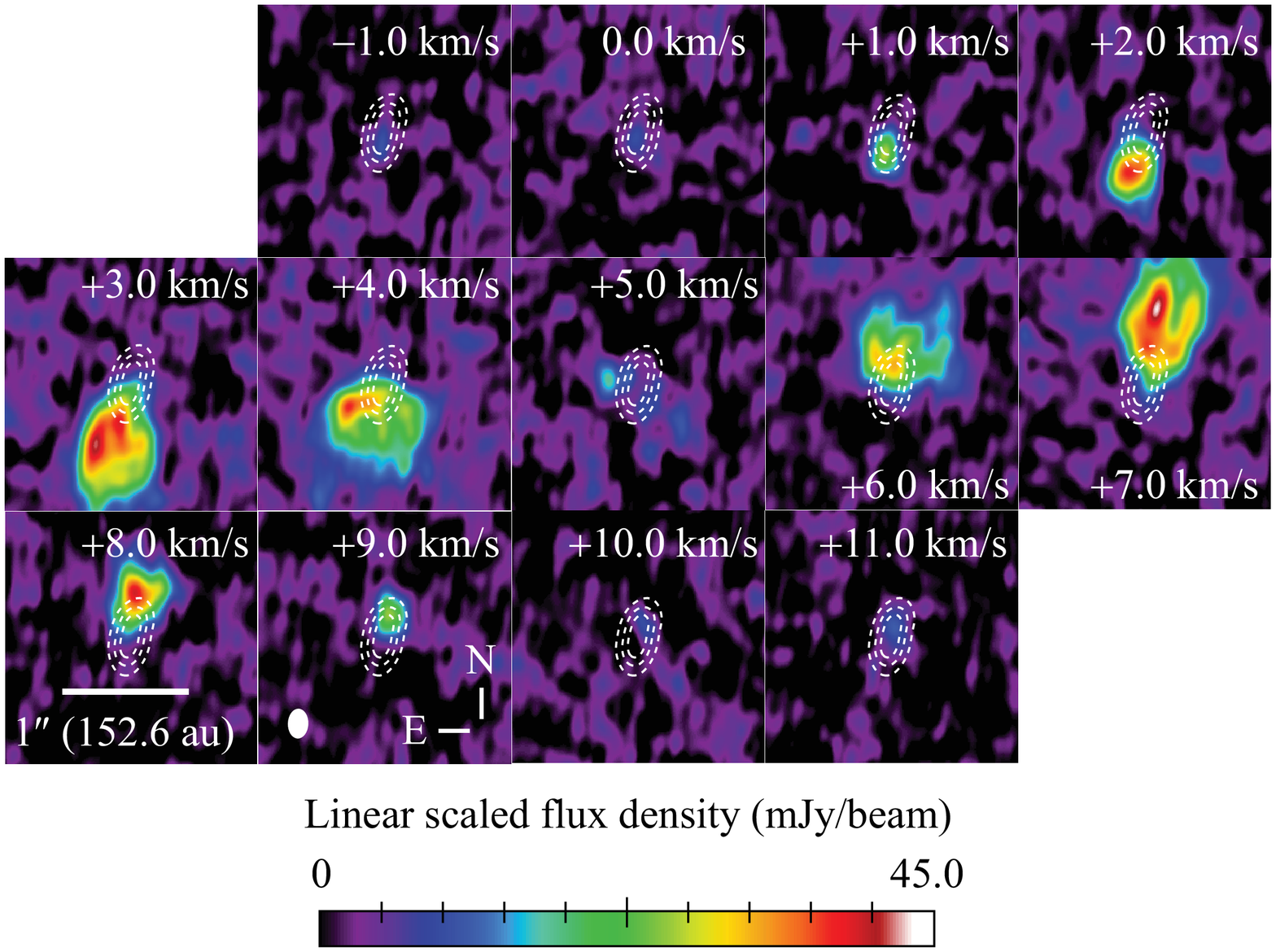}
\end{centering}
\caption{Channel maps of $^{12}$CO~$J=2-1$. The r.m.s noise at the 1.0~km/s bin is 2.37~mJy/beam with a beam size of 235~$\times$~161~mas at a PA of $-$2.0$^{\circ}$.}\label{figA1}
\end{figure}

Figure~\ref{figA1} shows the $^{12}$CO~$J=2-1$ channel maps at $-$1.0 to $+$11.0~km/s.

\begin{figure}[ht!]
\begin{centering}
\includegraphics[clip,width=\linewidth]{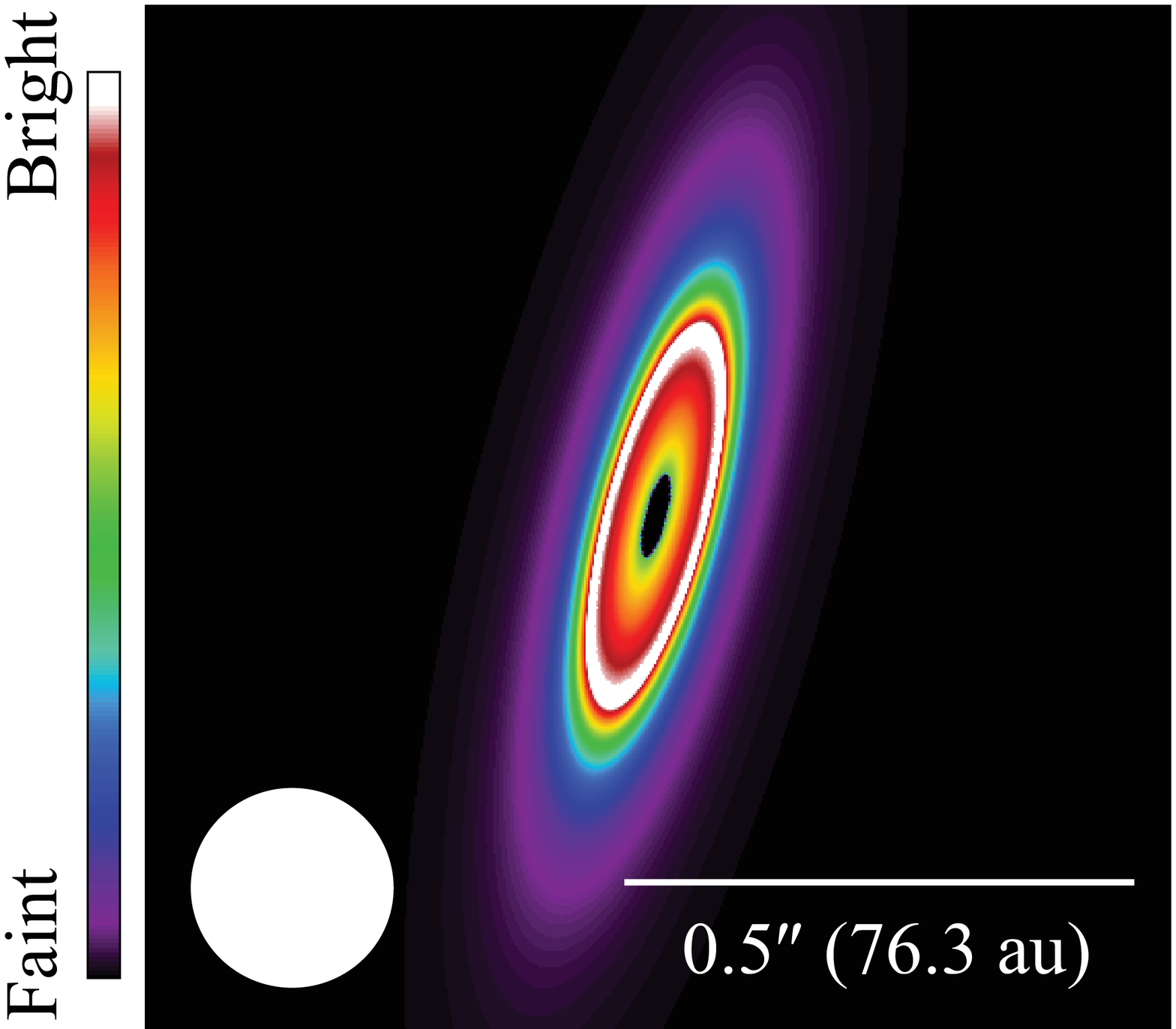}
\end{centering}
\caption{Best-fit model image at band~6 with row resolution. White circle represents the beam shape of 198~$\times$~195~mas at a PA of $-$88.5\degr in our ALMA observations.
}\label{figA3}
\end{figure}

Figure~\ref{figA3} shows the best-fit model image of the disk around Sz~84 based on MCMC model fitting in \S~\ref{subsec:mcmc}.

\begin{figure}[ht!]
\begin{centering}
\includegraphics[clip,width=\linewidth]{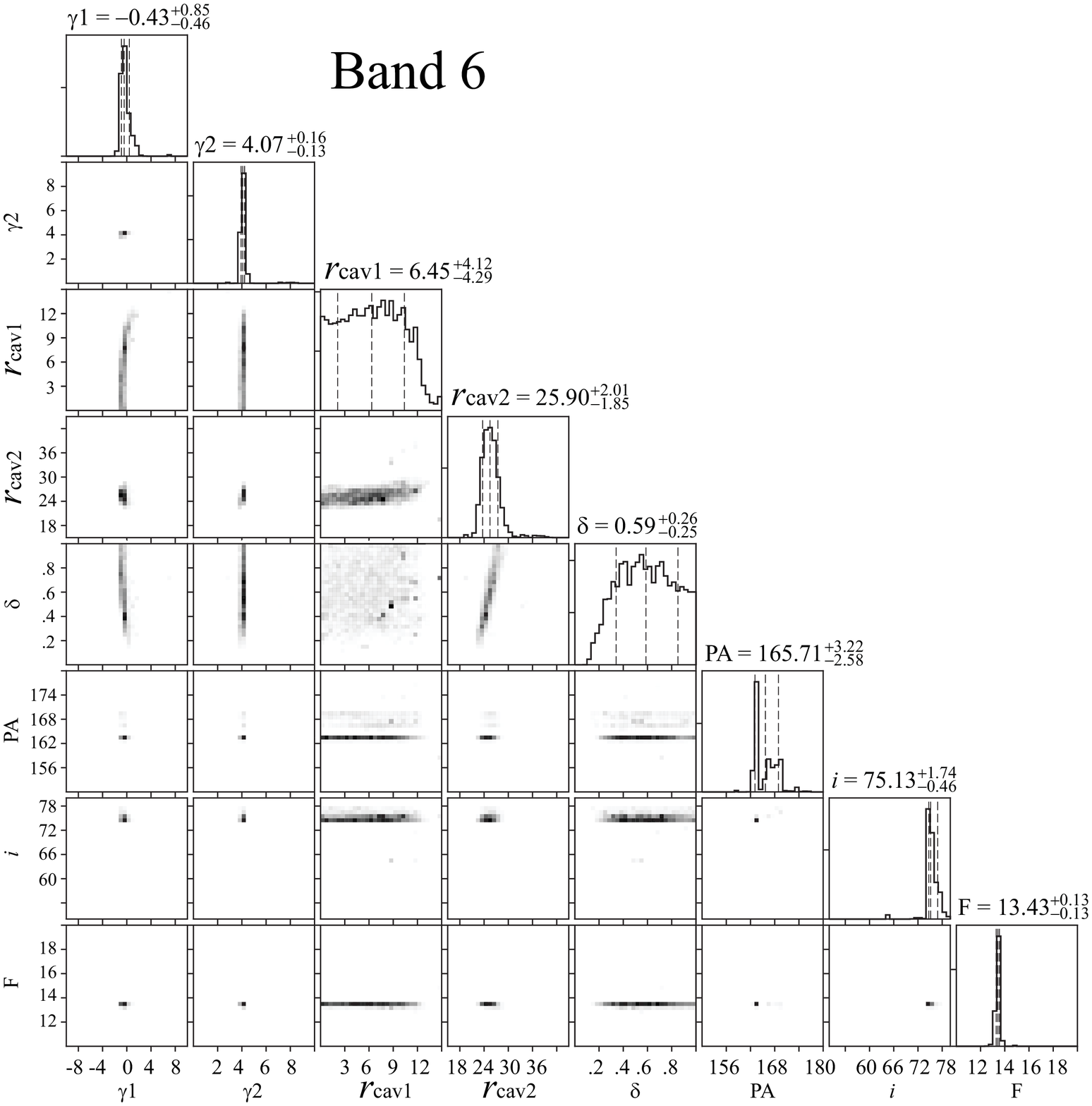}
\end{centering}
\caption{Corner plot of the MCMC posteriors calculated in the visibility fitting for band~6 data in \S~\ref{subsec:mcmc}. The histograms on the diagonal are marginal distributions of 8 free parameters. The ranges of parameters are described in Table~\ref{tab:table1}. The vertical dashed lines in the histograms represent the median values and the 1~$\sigma$ confidence intervals of parameters computed from the 16th and 84th percentiles. The off-diagonal plots show the correlation for corresponding pairs of parameters.}\label{figA2}
\end{figure}

\begin{figure}[ht!]
\begin{centering}
\includegraphics[clip,width=\linewidth]{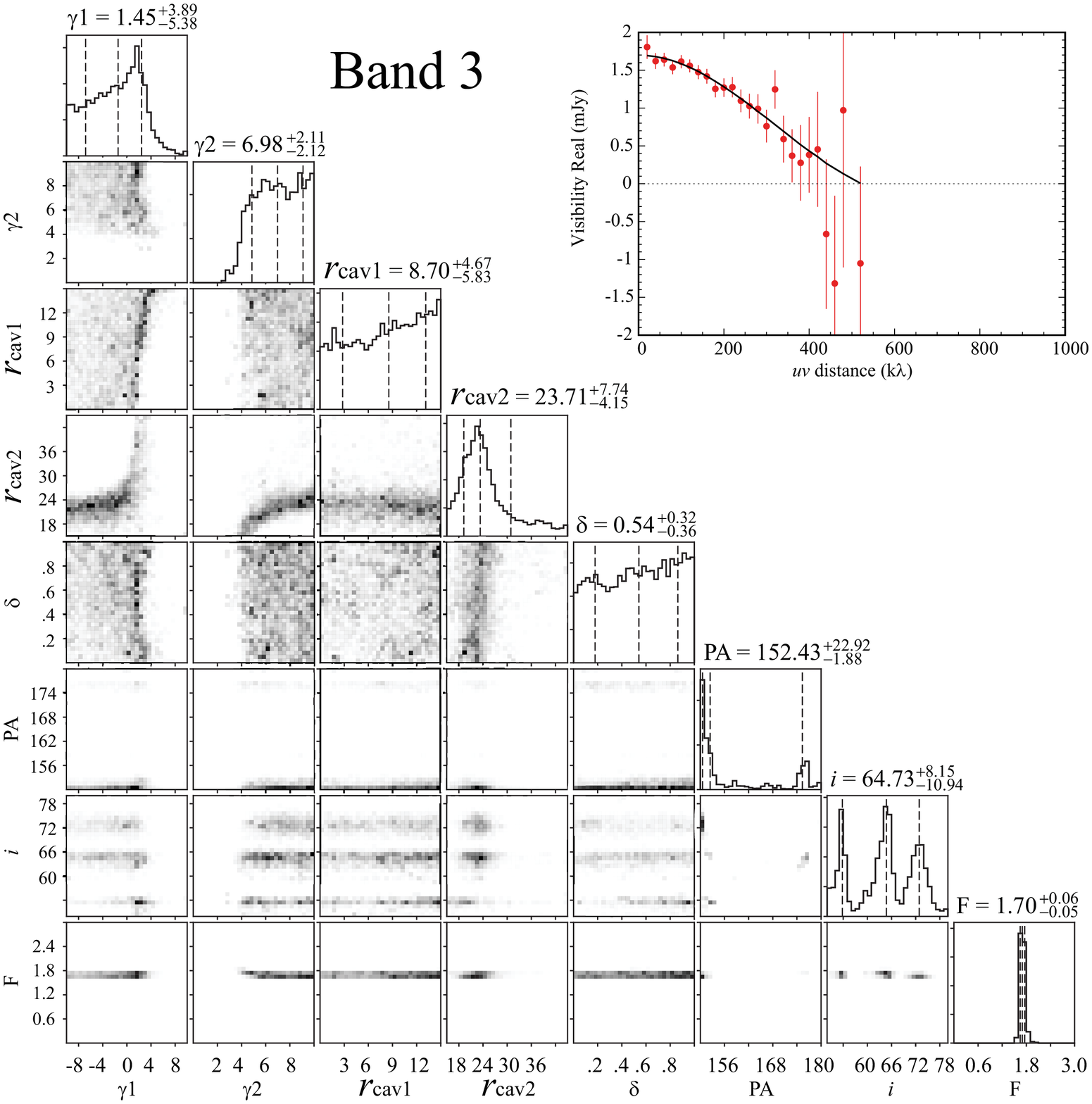}
\end{centering}
\caption{Same as Figure~\ref{figA2} but for band~3 data. The top-right panel is the real part of visibilities in observations (red dots) and the best-fit model (black line).}\label{figA5}
\end{figure}

\begin{figure}[ht!]
\begin{centering}
\includegraphics[clip,width=\linewidth]{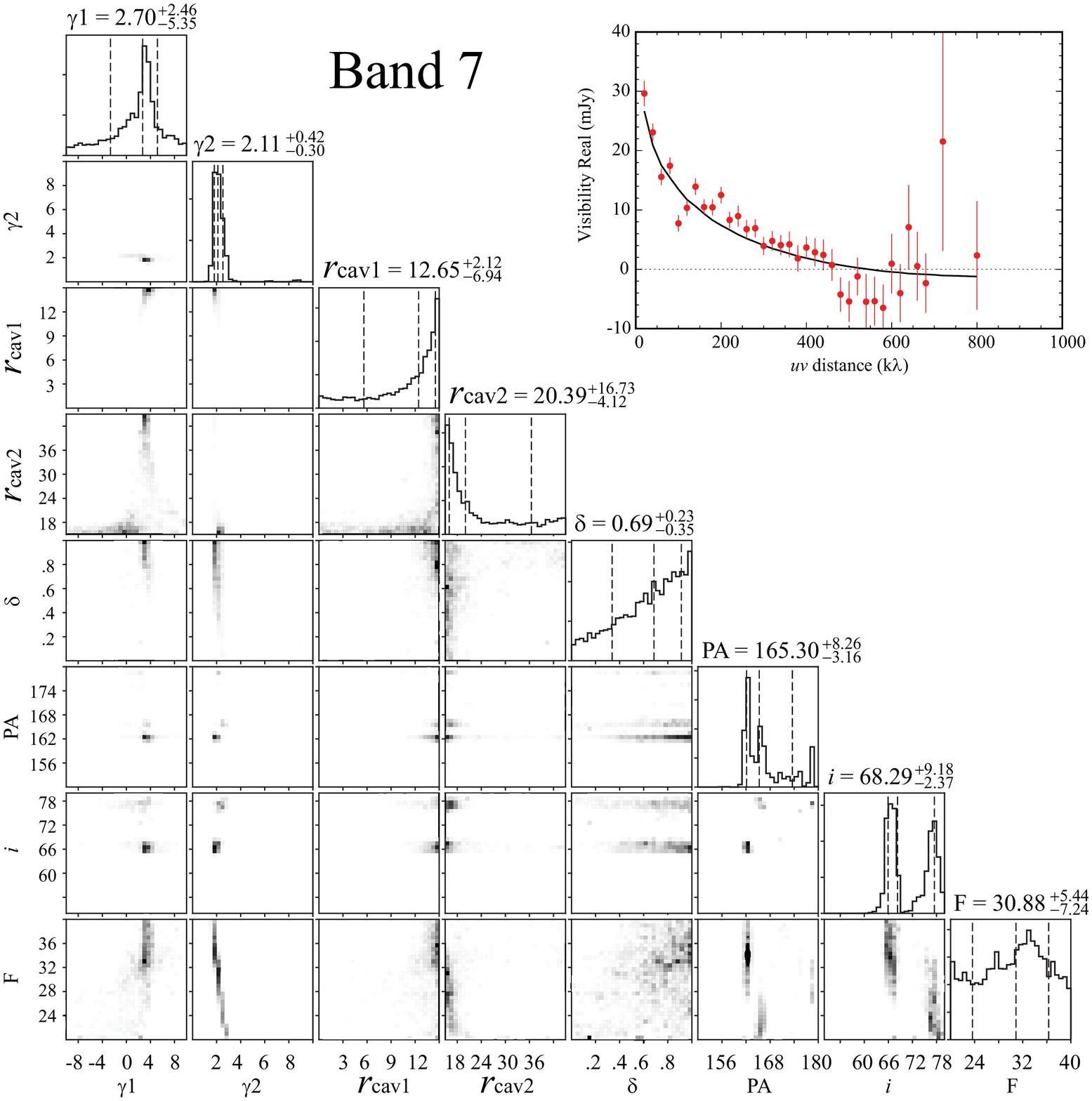}
\end{centering}
\caption{Same as Figure~\ref{figA2} but for band~7 data. The top-right panel is the real part of visibilities in observations (red dots) and the best-fit model (black line).}\label{figA6}
\end{figure}

Figures~\ref{figA2}, \ref{figA5}, and \ref{figA6} show corner plots of the MCMC posteriors calculated in the visibility fitting for bands~6, 3, and 7, respectively, in \S~\ref{subsec:mcmc}.

\begin{deluxetable*}{ccccccccc}
\tabletypesize{\footnotesize}
\tablewidth{0pt} 
\tablenum{5}
\tablecaption{Results of MCMC fitting for band~3 and 7 data and their parameter ranges \label{tab:table4}}
\tablehead{
\colhead{Band} & \colhead{Flux} & \colhead{$r_{\rm cav1}$} & \colhead{$r_{\rm cav2}$} & \colhead{$\gamma_{1}$} & \colhead{$\gamma_{2}$} & \colhead{$\delta$} & \colhead{$i$} & \colhead{P.A.}  \\
\colhead{} & \colhead{(mJy)} & \colhead{(au)}              & \colhead{(au)}  & \colhead{}   & \colhead{} &\colhead{}& \colhead{($\degr$)} & \colhead{($\degr$)} 
}
\colnumbers
\startdata
Band 3 & 1.70$^{+0.06}_{-0.05}$ & 8.70$^{+4.67}_{-5.83}$ & 23.71$^{+7.74}_{-4.15}$ & 1.45$^{+3.89}_{-5.38}$ & 6.98$^{+2.11}_{-2.12}$ & 0.54$^{+0.32}_{-0.36}$ & 64.73$^{+8.15}_{-10.94}$ & 152.43$^{+22.92}_{-1.88}$     \\
&\{0.0-3.0\} & \{0.0 .. 21.4\} & \{15.3 .. 45.8\} & \{0.0 .. 10\} & \{-10 .. 10\} & \{0.001 .. 1.0\} & \{50 .. 80\} & \{150 .. 180\}  \\
\hline 
Band 7 & 30.88$^{+5.44}_{-7.24}$ & 12.65$^{+2.12}_{-6.94}$ & 20.39$^{+16.73}_{-4.12}$ & 2.70$^{+2.46}_{-5.35}$ & 2.11$^{+0.42}_{-0.30}$ & 0.69$^{+0.23}_{-0.35}$ & 68.29$^{+9.18}_{-2.37}$ & 165.30$^{+8.26}_{-3.16}$     \\
&\{20.0-40.0\} & \{0.0 .. 21.4\} & \{15.3 .. 45.8\} & \{0.0 .. 10\} & \{-10 .. 10\} & \{0.001 .. 1.0\} & \{50 .. 80\} & \{150 .. 180\}  \\
\hline
\enddata
\tablecomments{
 Values in parentheses are parameter ranges in our MCMC calculations. }
\end{deluxetable*}

Table~\ref{tab:phot} shows photometric and spectroscopic data of Sz~84 used in SED fitting in \S~\ref{subsec:mcrt}.

\begin{deluxetable*}{lll}
\tabletypesize{\scriptsize}
\tablewidth{0pt} 
\tablenum{5}
\tablecaption{Photometric and spectroscopic values of Sz~84\label{tab:phot}}
\tablehead{
\colhead{Wavelength} & \colhead{$\lambda F_{\lambda}$} & \colhead{Reference}  \\
\colhead{($\mu$m)}   & \colhead{(10$^{-13}$ erg s$^{-1}$ cm$^{-2}$)} & \colhead{}  \\
}
\startdata
0.44         & 132.0                 & USNO-B1.0 \citep{mone2003} \\
0.64         & 400.8                 & USNO-B1.0 \citep{mone2003} \\
0.80         & 1388.1                & USNO-B1.0 \citep{mone2003} \\
1.235        & 2042.6 $\pm$ 43.2     & 2MASS \citep{cutr2003a} \\
1.662        & 1758.1 $\pm$ 37.2     & 2MASS \citep{cutr2003a} \\
2.159        & 1152.8 $\pm$ 24.4     & 2MASS \citep{cutr2003a}\\
3.350        & 37.12 $\pm$ 7.9       & WISE  \citep{cutr2014a}\\
4.600        & 20.80 $\pm$ 4.0       & WISE  \citep{cutr2014a} \\
5.308        & 110.1 $\pm$ 8.3 & Spitzer/IRS (Spitzer Heritage Archive)\\
5.520        & 104.7 $\pm$ 6.8 & Spitzer/IRS (Spitzer Heritage Archive)\\
5.731        & 94.0  $\pm$ 6.4 & Spitzer/IRS (Spitzer Heritage Archive)\\
5.943        & 94.2 $\pm$ 6.2 & Spitzer/IRS (Spitzer Heritage Archive)\\
6.155        & 90.5 $\pm$ 6.0 & Spitzer/IRS (Spitzer Heritage Archive)\\
6.366        & 79.6 $\pm$ 6.4 & Spitzer/IRS (Spitzer Heritage Archive)\\
6.578        & 64.4 $\pm$ 7.1 & Spitzer/IRS (Spitzer Heritage Archive)\\
6.790        & 63.7 $\pm$ 7.6 & Spitzer/IRS (Spitzer Heritage Archive)\\
7.002        & 57.7 $\pm$ 8.3 & Spitzer/IRS (Spitzer Heritage Archive)\\
7.213        & 55.0 $\pm$ 8.2 & Spitzer/IRS (Spitzer Heritage Archive)\\
7.425        & 53.3 $\pm$ 8.9 & Spitzer/IRS (Spitzer Heritage Archive)\\
7.701        & 37.2 $\pm$ 4.9 & Spitzer/IRS (Spitzer Heritage Archive)\\
8.120        & 37.3 $\pm$ 3.8 & Spitzer/IRS (Spitzer Heritage Archive)\\
8.544        & 40.1 $\pm$ 3.4 & Spitzer/IRS (Spitzer Heritage Archive)\\
8.967        & 31.2 $\pm$ 3.3 & Spitzer/IRS (Spitzer Heritage Archive)\\
9.391        & 27.1 $\pm$ 3.2 & Spitzer/IRS (Spitzer Heritage Archive)\\
9.814        & 28.1 $\pm$ 2.7 & Spitzer/IRS (Spitzer Heritage Archive)\\
10.24        & 24.7 $\pm$ 2.4 & Spitzer/IRS (Spitzer Heritage Archive)\\
10.66        & 23.7 $\pm$ 2.6 & Spitzer/IRS (Spitzer Heritage Archive)\\
11.08        & 18.1 $\pm$ 3.0 & Spitzer/IRS (Spitzer Heritage Archive)\\
11.51        & 14.3 $\pm$ 3.1 & Spitzer/IRS (Spitzer Heritage Archive)\\
11.93        & 15.5 $\pm$ 3.2 & Spitzer/IRS (Spitzer Heritage Archive)\\
12.35        & 10.9 $\pm$ 3.0 & Spitzer/IRS (Spitzer Heritage Archive)\\
12.78        & 14.7 $\pm$ 2.9 & Spitzer/IRS (Spitzer Heritage Archive)\\
13.20        & 11.5 $\pm$ 3.5 & Spitzer/IRS (Spitzer Heritage Archive)\\
13.62        & 11.1 $\pm$ 3.8 & Spitzer/IRS (Spitzer Heritage Archive)\\
14.05        & 10.9 $\pm$ 4.3 & Spitzer/IRS (Spitzer Heritage Archive)\\
14.58        & 6.7  $\pm$ 2.5 & Spitzer/IRS (Spitzer Heritage Archive)\\
15.17        & 7.4  $\pm$ 2.8 & Spitzer/IRS (Spitzer Heritage Archive)\\
15.76        & 11.4 $\pm$ 2.2 & Spitzer/IRS (Spitzer Heritage Archive)\\
16.36        & 8.0  $\pm$ 3.0 & Spitzer/IRS (Spitzer Heritage Archive)\\
16.95        & 7.4  $\pm$ 2.8 & Spitzer/IRS (Spitzer Heritage Archive)\\
17.54        & 10.9 $\pm$ 2.7 & Spitzer/IRS (Spitzer Heritage Archive)\\
18.13        & 7.6  $\pm$ 3.3 & Spitzer/IRS (Spitzer Heritage Archive)\\
18.73        & 11.1 $\pm$ 3.2 & Spitzer/IRS (Spitzer Heritage Archive)\\
19.32        & 12.7 $\pm$ 4.7 & Spitzer/IRS (Spitzer Heritage Archive)\\
19.91        & 14.8 $\pm$ 3.8 & Spitzer/IRS (Spitzer Heritage Archive)\\
20.54        & 11.8 $\pm$ 4.6 & Spitzer/IRS (Spitzer Heritage Archive)\\
21.61        & 23.0 $\pm$ 3.2 & Spitzer/IRS (Spitzer Heritage Archive)\\
22.79        & 27.1 $\pm$ 2.4 & Spitzer/IRS (Spitzer Heritage Archive)\\
23.98        & 31.3 $\pm$ 2.2 & Spitzer/IRS (Spitzer Heritage Archive)\\
25.16        & 37.7 $\pm$ 2.3 & Spitzer/IRS (Spitzer Heritage Archive) \\
26.35        & 41.3 $\pm$ 2.2 & Spitzer/IRS (Spitzer Heritage Archive)\\
27.53        & 49.1 $\pm$ 2.1 & Spitzer/IRS (Spitzer Heritage Archive)\\
28.72        & 57.6 $\pm$ 2.6 & Spitzer/IRS (Spitzer Heritage Archive)\\
29.90        & 61.4 $\pm$ 2.4 & Spitzer/IRS (Spitzer Heritage Archive)\\
31.09        & 67.6 $\pm$ 3.1 & Spitzer/IRS (Spitzer Heritage Archive)\\
32.27        & 78.9 $\pm$ 3.4 & Spitzer/IRS (Spitzer Heritage Archive)\\
33.46        & 87.7 $\pm$ 5.2 & Spitzer/IRS (Spitzer Heritage Archive)\\
34.65        & 103.2 $\pm$ 4.9 & Spitzer/IRS (Spitzer Heritage Archive)\\
35.83        & 109.7 $\pm$ 9.6 & Spitzer/IRS (Spitzer Heritage Archive)\\
37.02        & 112.3 $\pm$ 9.2 & Spitzer/IRS (Spitzer Heritage Archive)\\
63           & 190.0 $\pm$ 4.8 & Herschel \citep{rivi2016a} \\
70           & 105.0 $\pm$ 3.1 & Spitzer/MIPS \citep{meri2010a}\\
909          & 1.14  $\pm$ 0.14  & ALMA \citep{vandermarel18a}\\
1300         & 0.30  $\pm$ 0.003 & ALMA (this work) \\
\enddata
\tablecomments{
The value of $A_{V}=$0.5~mag \citep{mana14} is applied here.}
\end{deluxetable*}

\begin{figure}[ht!]
\begin{centering}
\includegraphics[clip,width=\linewidth]{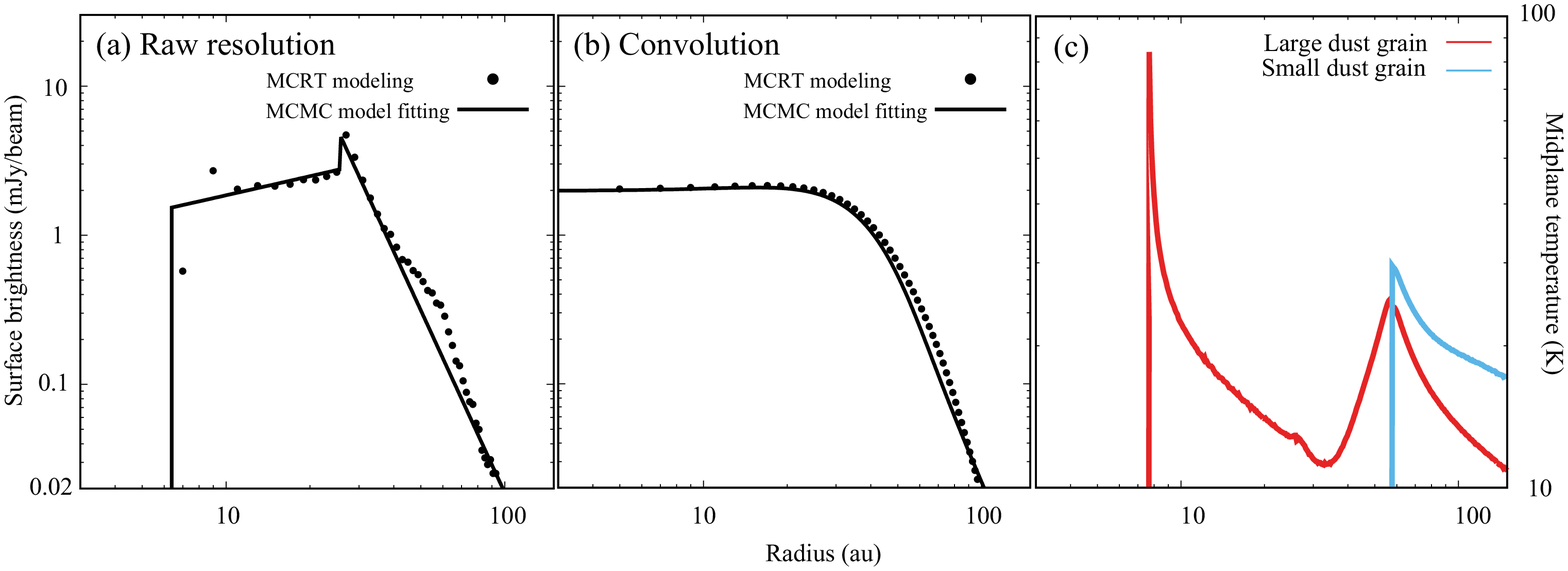}
\end{centering}
\caption{Radial profiles of modeled dust continuum images at (a) raw resolution and (b) convolution with the ALMA beam shape of 0\farcs2 (30~au) based on MCMC model fitting and MCRT modeling. (c) Midplane temperature of large and small dust grains in the fiducial model calculated based on MCRT modeling in \S~\ref{subsec:mcrt}. The maximum deviation at $r\sim$50~au in panel (b) is $\sim$119~$\mu$Jy (1.9~$\sigma$).
}\label{figA4}
\end{figure}

Figure~\ref{figA4} shows the azimuthally averaged radial profile at $r<$150~au based on MCMC model fitting and MCRT modeling in \S~\ref{subsec:mcmc} and \S~\ref{subsec:mcrt}, respectively, to test the consistency of the surface brightness in the two models.

\bibliography{z.do_not_edit_20200916}{}

\begin{thebibliography}{}
\expandafter\ifx\csname natexlab\endcsname\relax\def\natexlab#1{#1}\fi
\providecommand{\url}[1]{\href{#1}{#1}}
\providecommand{\dodoi}[1]{doi:~\href{http://doi.org/#1}{\nolinkurl{#1}}}
\providecommand{\doeprint}[1]{\href{http://ascl.net/#1}{\nolinkurl{http://ascl.net/#1}}}
\providecommand{\doarXiv}[1]{\href{https://arxiv.org/abs/#1}{\nolinkurl{https://arxiv.org/abs/#1}}}

\bibitem[{{Alcal{\'a}} {et~al.}(2014){Alcal{\'a}}, {Natta}, {Manara}, {Spezzi},
  {Stelzer}, {Frasca}, {Biazzo}, {Covino}, {Randich}, {Rigliaco}, {Testi},
  {Comer{\'o}n}, {Cupani}, \& {D'Elia}}]{alca2014}
{Alcal{\'a}}, J.~M., {Natta}, A., {Manara}, C.~F., {et~al.} 2014, \aap, 561,
  A2, \dodoi{10.1051/0004-6361/201322254}

\bibitem[{{Alcal{\'a}} {et~al.}(2017){Alcal{\'a}}, {Manara}, {Natta}, {Frasca},
  {Testi}, {Nisini}, {Stelzer}, {Williams}, {Antoniucci}, {Biazzo}, {Covino},
  {Esposito}, {Getman}, \& {Rigliaco}}]{alca2017}
{Alcal{\'a}}, J.~M., {Manara}, C.~F., {Natta}, A., {et~al.} 2017, \aap, 600,
  A20, \dodoi{10.1051/0004-6361/201629929}

\bibitem[{{Alexander} {et~al.}(2014){Alexander}, {Pascucci}, {Andrews},
  {Armitage}, \& {Cieza}}]{alex2014}
{Alexander}, R., {Pascucci}, I., {Andrews}, S., {Armitage}, P., \& {Cieza}, L.
  2014, in Protostars and Planets VI, ed. H.~{Beuther}, R.~S. {Klessen}, C.~P.
  {Dullemond}, \& T.~{Henning}, 475,
  \dodoi{10.2458/azu_uapress_9780816531240-ch021}

\bibitem[{{Andrews} {et~al.}(2013){Andrews}, {Rosenfeld}, {Kraus}, \&
  {Wilner}}]{andr2013}
{Andrews}, S.~M., {Rosenfeld}, K.~A., {Kraus}, A.~L., \& {Wilner}, D.~J. 2013,
  \apj, 771, 129, \dodoi{10.1088/0004-637X/771/2/129}

\bibitem[{{Ansdell} {et~al.}(2016){Ansdell}, {Williams}, {van der Marel},
  {Carpenter}, {Guidi}, {Hogerheijde}, {Mathews}, {Manara}, {Miotello},
  {Natta}, {Oliveira}, {Tazzari}, {Testi}, {van Dishoeck}, \& {van
  Terwisga}}]{ansd2016a}
{Ansdell}, M., {Williams}, J.~P., {van der Marel}, N., {et~al.} 2016, \apj,
  828, 46, \dodoi{10.3847/0004-637X/828/1/46}

\bibitem[{{Ansdell} {et~al.}(2018){Ansdell}, {Williams}, {Trapman}, {van
  Terwisga}, {Facchini}, {Manara}, {van der Marel}, {Miotello}, {Tazzari},
  {Hogerheijde}, {Guidi}, {Testi}, \& {van Dishoeck}}]{ansd2018a}
{Ansdell}, M., {Williams}, J.~P., {Trapman}, L., {et~al.} 2018, \apj, 859, 21,
  \dodoi{10.3847/1538-4357/aab890}

\bibitem[{{Beichman} {et~al.}(2005){Beichman}, {Bryden}, {Gautier},
  {Stapelfeldt}, {Werner}, {Misselt}, {Rieke}, {Stansberry}, \&
  {Trilling}}]{beichman05}
{Beichman}, C.~A., {Bryden}, G., {Gautier}, T.~N., {et~al.} 2005, \apj, 626,
  1061, \dodoi{10.1086/430059}

\bibitem[{{Birnstiel} {et~al.}(2012){Birnstiel}, {Andrews}, \&
  {Ercolano}}]{birn2012a}
{Birnstiel}, T., {Andrews}, S.~M., \& {Ercolano}, B. 2012, \aap, 544, A79,
  \dodoi{10.1051/0004-6361/201219262}

\bibitem[{{Brauer} {et~al.}(2008){Brauer}, {Dullemond}, \&
  {Henning}}]{brauer+2008}
{Brauer}, F., {Dullemond}, C.~P., \& {Henning}, T. 2008, \aap, 480, 859,
  \dodoi{10.1051/0004-6361:20077759}

\bibitem[{{Clarke} {et~al.}(2001){Clarke}, {Gendrin}, \&
  {Sotomayor}}]{clar2001a}
{Clarke}, C.~J., {Gendrin}, A., \& {Sotomayor}, M. 2001, \mnras, 328, 485,
  \dodoi{10.1046/j.1365-8711.2001.04891.x}

\bibitem[{{Cutri} \& {et al.}(2014)}]{cutr2014a}
{Cutri}, R.~M., \& {et al.} 2014, VizieR Online Data Catalog, II/328

\bibitem[{{Cutri} {et~al.}(2003){Cutri}, {Skrutskie}, {van Dyk}, {Beichman},
  {Carpenter}, {Chester}, {Cambresy}, {Evans}, {Fowler}, {Gizis}, {Howard},
  {Huchra}, {Jarrett}, {Kopan}, {Kirkpatrick}, {Light}, {Marsh}, {McCallon},
  {Schneider}, {Stiening}, {Sykes}, {Weinberg}, {Wheaton}, {Wheelock}, \&
  {Zacarias}}]{cutr2003a}
{Cutri}, R.~M., {Skrutskie}, M.~F., {van Dyk}, S., {et~al.} 2003, VizieR Online
  Data Catalog, II/246

\bibitem[{{D'Alessio} {et~al.}(2006){D'Alessio}, {Calvet}, {Hartmann},
  {Franco-Hern{\'a}ndez}, \& {Serv{\'\i}n}}]{dale2006}
{D'Alessio}, P., {Calvet}, N., {Hartmann}, L., {Franco-Hern{\'a}ndez}, R., \&
  {Serv{\'\i}n}, H. 2006, \apj, 638, 314, \dodoi{10.1086/498861}

\bibitem[{{de Juan Ovelar} {et~al.}(2013){de Juan Ovelar}, {Min}, {Dominik},
  {Thalmann}, {Pinilla}, {Benisty}, \& {Birnstiel}}]{dujuanovelar2013}
{de Juan Ovelar}, M., {Min}, M., {Dominik}, C., {et~al.} 2013, \aap, 560, A111,
  \dodoi{10.1051/0004-6361/201322218}

\bibitem[{{Dong} {et~al.}(2012){Dong}, {Rafikov}, {Zhu}, {Hartmann}, {Whitney},
  {Brandt}, {Muto}, {Hashimoto}, {Grady}, {Follette}, {Kuzuhara}, {Tanii},
  {Itoh}, {Thalmann}, {Wisniewski}, {Mayama}, {Janson}, {Abe}, {Brandner},
  {Carson}, {Egner}, {Feldt}, {Goto}, {Guyon}, {Hayano}, {Hayashi}, {Hayashi},
  {Henning}, {Hodapp}, {Honda}, {Inutsuka}, {Ishii}, {Iye}, {Kandori}, {Knapp},
  {Kudo}, {Kusakabe}, {Matsuo}, {McElwain}, {Miyama}, {Morino}, {Moro-Martin},
  {Nishimura}, {Pyo}, {Suto}, {Suzuki}, {Takami}, {Takato}, {Terada}, {Tomono},
  {Turner}, {Watanabe}, {Yamada}, {Takami}, {Usuda}, \&
  {Tamura}}]{dong12cavity}
{Dong}, R., {Rafikov}, R., {Zhu}, Z., {et~al.} 2012, \apj, 750, 161,
  \dodoi{10.1088/0004-637X/750/2/161}

\bibitem[{{Draine}(2006)}]{drai2006}
{Draine}, B.~T. 2006, \apj, 636, 1114, \dodoi{10.1086/498130}

\bibitem[{{Dullemond} \& {Dominik}(2005)}]{dull2005a}
{Dullemond}, C.~P., \& {Dominik}, C. 2005, \aap, 434, 971,
  \dodoi{10.1051/0004-6361:20042080}

\bibitem[{{Ercolano} \& {Pascucci}(2017)}]{erco2017a}
{Ercolano}, B., \& {Pascucci}, I. 2017, Royal Society Open Science, 4, 170114,
  \dodoi{10.1098/rsos.170114}

\bibitem[{{Espaillat} {et~al.}(2010){Espaillat}, {D'Alessio}, {Hern{\'a}ndez},
  {Nagel}, {Luhman}, {Watson}, {Calvet}, {Muzerolle}, \&
  {McClure}}]{espaillat10}
{Espaillat}, C., {D'Alessio}, P., {Hern{\'a}ndez}, J., {et~al.} 2010, \apj,
  717, 441, \dodoi{10.1088/0004-637X/717/1/441}

\bibitem[{{Espaillat} {et~al.}(2014){Espaillat}, {Muzerolle}, {Najita},
  {Andrews}, {Zhu}, {Calvet}, {Kraus}, {Hashimoto}, {Kraus}, \&
  {D'Alessio}}]{espaillat+14}
{Espaillat}, C., {Muzerolle}, J., {Najita}, J., {et~al.} 2014, Protostars and
  Planets VI, 497, \dodoi{10.2458/azu_uapress_9780816531240-ch022}

\bibitem[{{Facchini} {et~al.}(2018){Facchini}, {Pinilla}, {van Dishoeck}, \&
  {de Juan Ovelar}}]{facc2018b}
{Facchini}, S., {Pinilla}, P., {van Dishoeck}, E.~F., \& {de Juan Ovelar}, M.
  2018, \aap, 612, A104, \dodoi{10.1051/0004-6361/201731390}

\bibitem[{{Foreman-Mackey} {et~al.}(2013){Foreman-Mackey}, {Hogg}, {Lang}, \&
  {Goodman}}]{foreman-mackey+2013}
{Foreman-Mackey}, D., {Hogg}, D.~W., {Lang}, D., \& {Goodman}, J. 2013, \pasp,
  125, 306, \dodoi{10.1086/670067}

\bibitem[{{Francis} \& {van der Marel}(2020)}]{fran2020a}
{Francis}, L., \& {van der Marel}, N. 2020, arXiv e-prints, arXiv:2003.00079.
\newblock \doarXiv{2003.00079}

\bibitem[{{Gaia Collaboration} {et~al.}(2018){Gaia Collaboration}, {Brown},
  {Vallenari}, {Prusti}, {de Bruijne}, {Babusiaux}, {Bailer-Jones}, {Biermann},
  {Evans}, {Eyer}, {Jansen}, {Jordi}, {Klioner}, {Lammers}, {Lindegren},
  {Luri}, {Mignard}, {Panem}, {Pourbaix}, {Randich}, {Sartoretti}, {Siddiqui},
  {Soubiran}, {van Leeuwen}, {Walton}, {Arenou}, {Bastian}, {Cropper},
  {Drimmel}, {Katz}, {Lattanzi}, {Bakker}, {Cacciari}, {Casta{\~n}eda},
  {Chaoul}, {Cheek}, {De Angeli}, {Fabricius}, {Guerra}, {Holl}, {Masana},
  {Messineo}, {Mowlavi}, {Nienartowicz}, {Panuzzo}, {Portell}, {Riello},
  {Seabroke}, {Tanga}, {Th{\'e}venin}, {Gracia-Abril}, {Comoretto},
  {Garcia-Reinaldos}, {Teyssier}, {Altmann}, {Andrae}, {Audard},
  {Bellas-Velidis}, {Benson}, {Berthier}, {Blomme}, {Burgess}, {Busso},
  {Carry}, {Cellino}, {Clementini}, {Clotet}, {Creevey}, {Davidson}, {De
  Ridder}, {Delchambre}, {Dell'Oro}, {Ducourant},
  {Fern{\'a}ndez-Hern{\'a}ndez}, {Fouesneau}, {Fr{\'e}mat}, {Galluccio},
  {Garc{\'\i}a-Torres}, {Gonz{\'a}lez-N{\'u}{\~n}ez}, {Gonz{\'a}lez-Vidal},
  {Gosset}, {Guy}, {Halbwachs}, {Hambly}, {Harrison}, {Hern{\'a}ndez},
  {Hestroffer}, {Hodgkin}, {Hutton}, {Jasniewicz}, {Jean-Antoine-Piccolo},
  {Jordan}, {Korn}, {Krone-Martins}, {Lanzafame}, {Lebzelter}, {L{\"o}ffler},
  {Manteiga}, {Marrese}, {Mart{\'\i}n-Fleitas}, {Moitinho}, {Mora}, {Muinonen},
  {Osinde}, {Pancino}, {Pauwels}, {Petit}, {Recio-Blanco}, {Richards},
  {Rimoldini}, {Robin}, {Sarro}, {Siopis}, {Smith}, {Sozzetti}, {S{\"u}veges},
  {Torra}, {van Reeven}, {Abbas}, {Abreu Aramburu}, {Accart}, {Aerts},
  {Altavilla}, {{\'A}lvarez}, {Alvarez}, {Alves}, {Anderson}, {Andrei},
  {Anglada Varela}, {Antiche}, {Antoja}, {Arcay}, {Astraatmadja}, {Bach},
  {Baker}, {Balaguer-N{\'u}{\~n}ez}, {Balm}, {Barache}, {Barata}, {Barbato},
  {Barblan}, {Barklem}, {Barrado}, {Barros}, {Barstow}, {Bartholom{\'e}
  Mu{\~n}oz}, {Bassilana}, {Becciani}, {Bellazzini}, {Berihuete}, {Bertone},
  {Bianchi}, {Bienaym{\'e}}, {Blanco-Cuaresma}, {Boch}, {Boeche}, {Bombrun},
  {Borrachero}, {Bossini}, {Bouquillon}, {Bourda}, {Bragaglia}, {Bramante},
  {Breddels}, {Bressan}, {Brouillet}, {Br{\"u}semeister}, {Brugaletta},
  {Bucciarelli}, {Burlacu}, {Busonero}, {Butkevich}, {Buzzi}, {Caffau},
  {Cancelliere}, {Cannizzaro}, {Cantat-Gaudin}, {Carballo}, {Carlucci},
  {Carrasco}, {Casamiquela}, {Castellani}, {Castro-Ginard}, {Charlot},
  {Chemin}, {Chiavassa}, {Cocozza}, {Costigan}, {Cowell}, {Crifo}, {Crosta},
  {Crowley}, {Cuypers}, {Dafonte}, {Damerdji}, {Dapergolas}, {David}, {David},
  {de Laverny}, {De Luise}, {De March}, {de Martino}, {de Souza}, {de Torres},
  {Debosscher}, {del Pozo}, {Delbo}, {Delgado}, {Delgado}, {Di Matteo},
  {Diakite}, {Diener}, {Distefano}, {Dolding}, {Drazinos}, {Dur{\'a}n},
  {Edvardsson}, {Enke}, {Eriksson}, {Esquej}, {Eynard Bontemps}, {Fabre},
  {Fabrizio}, {Faigler}, {Falc{\~a}o}, {Farr{\`a}s Casas}, {Federici},
  {Fedorets}, {Fernique}, {Figueras}, {Filippi}, {Findeisen}, {Fonti},
  {Fraile}, {Fraser}, {Fr{\'e}zouls}, {Gai}, {Galleti}, {Garabato},
  {Garc{\'\i}a-Sedano}, {Garofalo}, {Garralda}, {Gavel}, {Gavras}, {Gerssen},
  {Geyer}, {Giacobbe}, {Gilmore}, {Girona}, {Giuffrida}, {Glass}, {Gomes},
  {Granvik}, {Gueguen}, {Guerrier}, {Guiraud}, {Guti{\'e}rrez-S{\'a}nchez},
  {Haigron}, {Hatzidimitriou}, {Hauser}, {Haywood}, {Heiter}, {Helmi}, {Heu},
  {Hilger}, {Hobbs}, {Hofmann}, {Holland}, {Huckle}, {Hypki}, {Icardi},
  {Jan{\ss}en}, {Jevardat de Fombelle}, {Jonker}, {Juh{\'a}sz}, {Julbe},
  {Karampelas}, {Kewley}, {Klar}, {Kochoska}, {Kohley}, {Kolenberg},
  {Kontizas}, {Kontizas}, {Koposov}, {Kordopatis}, {Kostrzewa-Rutkowska},
  {Koubsky}, {Lambert}, {Lanza}, {Lasne}, {Lavigne}, {Le Fustec}, {Le
  Poncin-Lafitte}, {Lebreton}, {Leccia}, {Leclerc}, {Lecoeur-Taibi},
  {Lenhardt}, {Leroux}, {Liao}, {Licata}, {Lindstr{\o}m}, {Lister}, {Livanou},
  {Lobel}, {L{\'o}pez}, {Managau}, {Mann}, {Mantelet}, {Marchal}, {Marchant},
  {Marconi}, {Marinoni}, {Marschalk{\'o}}, {Marshall}, {Martino}, {Marton},
  {Mary}, {Massari}, {Matijevi{\v{c}}}, {Mazeh}, {McMillan}, {Messina},
  {Michalik}, {Millar}, {Molina}, {Molinaro}, {Moln{\'a}r}, {Montegriffo},
  {Mor}, {Morbidelli}, {Morel}, {Morris}, {Mulone}, {Muraveva}, {Musella},
  {Nelemans}, {Nicastro}, {Noval}, {O'Mullane}, {Ord{\'e}novic},
  {Ord{\'o}{\~n}ez-Blanco}, {Osborne}, {Pagani}, {Pagano}, {Pailler},
  {Palacin}, {Palaversa}, {Panahi}, {Pawlak}, {Piersimoni}, {Pineau}, {Plachy},
  {Plum}, {Poggio}, {Poujoulet}, {Pr{\v{s}}a}, {Pulone}, {Racero}, {Ragaini},
  {Rambaux}, {Ramos-Lerate}, {Regibo}, {Reyl{\'e}}, {Riclet}, {Ripepi}, {Riva},
  {Rivard}, {Rixon}, {Roegiers}, {Roelens}, {Romero-G{\'o}mez}, {Rowell},
  {Royer}, {Ruiz-Dern}, {Sadowski}, {Sagrist{\`a} Sell{\'e}s}, {Sahlmann},
  {Salgado}, {Salguero}, {Sanna}, {Santana-Ros}, {Sarasso}, {Savietto},
  {Schultheis}, {Sciacca}, {Segol}, {Segovia}, {S{\'e}gransan}, {Shih},
  {Siltala}, {Silva}, {Smart}, {Smith}, {Solano}, {Solitro}, {Sordo}, {Soria
  Nieto}, {Souchay}, {Spagna}, {Spoto}, {Stampa}, {Steele},
  {Steidelm{\"u}ller}, {Stephenson}, {Stoev}, {Suess}, {Surdej}, {Szabados},
  {Szegedi-Elek}, {Tapiador}, {Taris}, {Tauran}, {Taylor}, {Teixeira},
  {Terrett}, {Teyssand ier}, {Thuillot}, {Titarenko}, {Torra Clotet}, {Turon},
  {Ulla}, {Utrilla}, {Uzzi}, {Vaillant}, {Valentini}, {Valette}, {van Elteren},
  {Van Hemelryck}, {van Leeuwen}, {Vaschetto}, {Vecchiato}, {Veljanoski},
  {Viala}, {Vicente}, {Vogt}, {von Essen}, {Voss}, {Votruba}, {Voutsinas},
  {Walmsley}, {Weiler}, {Wertz}, {Wevers}, {Wyrzykowski}, {Yoldas},
  {{\v{Z}}erjal}, {Ziaeepour}, {Zorec}, {Zschocke}, {Zucker}, {Zurbach}, \&
  {Zwitter}}]{gaia18}
{Gaia Collaboration}, {Brown}, A.~G.~A., {Vallenari}, A., {et~al.} 2018, \aap,
  616, A1, \dodoi{10.1051/0004-6361/201833051}

\bibitem[{{Garufi} {et~al.}(2013){Garufi}, {Quanz}, {Avenhaus}, {Buenzli},
  {Dominik}, {Meru}, {Meyer}, {Pinilla}, {Schmid}, \& {Wolf}}]{garu2013}
{Garufi}, A., {Quanz}, S.~P., {Avenhaus}, H., {et~al.} 2013, \aap, 560, A105,
  \dodoi{10.1051/0004-6361/201322429}

\bibitem[{{Haffert} {et~al.}(2019){Haffert}, {Bohn}, {de Boer}, {Snellen},
  {Brinchmann}, {Girard}, {Keller}, \& {Bacon}}]{haff19a}
{Haffert}, S.~Y., {Bohn}, A.~J., {de Boer}, J., {et~al.} 2019, Nature
  Astronomy, 329, \dodoi{10.1038/s41550-019-0780-5}

\bibitem[{{Hashimoto} {et~al.}(2015){Hashimoto}, {Tsukagoshi}, {Brown}, {Dong},
  {Muto}, {Zhu}, {Wisniewski}, {Ohashi}, {kudo}, {Kusakabe}, {Abe}, {Akiyama},
  {Brandner}, {Brandt}, {Carson}, {Currie}, {Egner}, {Feldt}, {Grady}, {Guyon},
  {Hayano}, {Hayashi}, {Hayashi}, {Henning}, {Hodapp}, {Ishii}, {Iye},
  {Janson}, {Kandori}, {Knapp}, {Kuzuhara}, {Kwon}, {Matsuo}, {McElwain},
  {Mayama}, {Mede}, {Miyama}, {Morino}, {Moro-Martin}, {Nishimura}, {Pyo},
  {Serabyn}, {Suenaga}, {Suto}, {Suzuki}, {Takahashi}, {Takami}, {Takato},
  {Terada}, {Thalmann}, {Tomono}, {Turner}, {Watanabe}, {Yamada}, {Takami},
  {Usuda}, \& {Tamura}}]{hashimoto+15}
{Hashimoto}, J., {Tsukagoshi}, T., {Brown}, J.~M., {et~al.} 2015, \apj, 799,
  43, \dodoi{10.1088/0004-637X/799/1/43}

\bibitem[{{Hayashi}(1981)}]{haya1981}
{Hayashi}, C. 1981, Progress of Theoretical Physics Supplement, 70, 35,
  \dodoi{10.1143/PTPS.70.35}

\bibitem[{{Hendler} {et~al.}(2020){Hendler}, {Pascucci}, {Pinilla}, {Tazzari},
  {Carpenter}, {Malhotra}, \& {Testi}}]{hend2020a}
{Hendler}, N., {Pascucci}, I., {Pinilla}, P., {et~al.} 2020, \apj, 895, 126,
  \dodoi{10.3847/1538-4357/ab70ba}

\bibitem[{{Johansen} {et~al.}(2014){Johansen}, {Blum}, {Tanaka}, {Ormel},
  {Bizzarro}, \& {Rickman}}]{joha2014a}
{Johansen}, A., {Blum}, J., {Tanaka}, H., {et~al.} 2014, in Protostars and
  Planets VI, ed. H.~{Beuther}, R.~S. {Klessen}, C.~P. {Dullemond}, \&
  T.~{Henning}, 547, \dodoi{10.2458/azu_uapress_9780816531240-ch024}

\bibitem[{{Kataoka} {et~al.}(2013){Kataoka}, {Tanaka}, {Okuzumi}, \&
  {Wada}}]{kata2013a}
{Kataoka}, A., {Tanaka}, H., {Okuzumi}, S., \& {Wada}, K. 2013, \aap, 557, L4,
  \dodoi{10.1051/0004-6361/201322151}

\bibitem[{{Keppler} {et~al.}(2018){Keppler}, {Benisty}, {M{\"u}ller},
  {Henning}, {van Boekel}, {Cantalloube}, {Ginski}, {van Holstein}, {Maire},
  {Pohl}, {Samland }, {Avenhaus}, {Baudino}, {Boccaletti}, {de Boer},
  {Bonnefoy}, {Chauvin}, {Desidera}, {Langlois}, {Lazzoni}, {Marleau},
  {Mordasini}, {Pawellek}, {Stolker}, {Vigan}, {Zurlo}, {Birnstiel},
  {Brandner}, {Feldt}, {Flock}, {Girard}, {Gratton}, {Hagelberg}, {Isella},
  {Janson}, {Juhasz}, {Kemmer}, {Kral}, {Lagrange}, {Launhardt}, {Matter},
  {M{\'e}nard}, {Milli}, {Molli{\`e}re}, {Olofsson}, {P{\'e}rez}, {Pinilla},
  {Pinte}, {Quanz}, {Schmidt}, {Udry}, {Wahhaj}, {Williams}, {Buenzli},
  {Cudel}, {Dominik}, {Galicher}, {Kasper}, {Lannier}, {Mesa}, {Mouillet},
  {Peretti}, {Perrot}, {Salter}, {Sissa}, {Wildi}, {Abe}, {Antichi},
  {Augereau}, {Baruffolo}, {Baudoz}, {Bazzon}, {Beuzit}, {Blanchard}, {Brems},
  {Buey}, {De Caprio}, {Carbillet}, {Carle}, {Cascone}, {Cheetham}, {Claudi},
  {Costille}, {Delboulb{\'e}}, {Dohlen}, {Fantinel}, {Feautrier}, {Fusco},
  {Giro}, {Gluck}, {Gry}, {Hubin}, {Hugot}, {Jaquet}, {Le Mignant}, {Llored},
  {Madec}, {Magnard}, {Martinez}, {Maurel}, {Meyer}, {M{\"o}ller-Nilsson},
  {Moulin}, {Mugnier}, {Orign{\'e}}, {Pavlov}, {Perret}, {Petit}, {Pragt},
  {Puget}, {Rabou}, {Ramos}, {Rigal}, {Rochat}, {Roelfsema}, {Rousset}, {Roux},
  {Salasnich}, {Sauvage}, {Sevin}, {Soenke}, {Stadler}, {Suarez}, {Turatto}, \&
  {Weber}}]{kepp18a}
{Keppler}, M., {Benisty}, M., {M{\"u}ller}, A., {et~al.} 2018, \aap, 617, A44,
  \dodoi{10.1051/0004-6361/201832957}

\bibitem[{{Kim} {et~al.}(1994){Kim}, {Martin}, \& {Hendry}}]{kim1994}
{Kim}, S.-H., {Martin}, P.~G., \& {Hendry}, P.~D. 1994, \apj, 422, 164,
  \dodoi{10.1086/173714}

\bibitem[{{Kley} \& {Nelson}(2012)}]{kley12}
{Kley}, W., \& {Nelson}, R.~P. 2012, \araa, 50, 211,
  \dodoi{10.1146/annurev-astro-081811-125523}

\bibitem[{{Krumholz} {et~al.}(2019){Krumholz}, {Ireland}, \&
  {Kratter}}]{krum2019a}
{Krumholz}, M.~R., {Ireland}, M.~J., \& {Kratter}, K.~M. 2019, arXiv e-prints,
  arXiv:1912.06788.
\newblock \doarXiv{1912.06788}

\bibitem[{{Kudo} {et~al.}(2018){Kudo}, {Hashimoto}, {Muto}, {Liu}, {Dong},
  {Hasegawa}, {Tsukagoshi}, \& {Konishi}}]{kudo2018}
{Kudo}, T., {Hashimoto}, J., {Muto}, T., {et~al.} 2018, \apjl, 868, L5,
  \dodoi{10.3847/2041-8213/aaeb1c}

\bibitem[{{Loomis} {et~al.}(2017){Loomis}, {{\"O}berg}, {Andrews}, \&
  {MacGregor}}]{loomis+17}
{Loomis}, R.~A., {{\"O}berg}, K.~I., {Andrews}, S.~M., \& {MacGregor}, M.~A.
  2017, \apj, 840, 23, \dodoi{10.3847/1538-4357/aa6c63}

\bibitem[{{Lynden-Bell} \& {Pringle}(1974)}]{lyndenbell74}
{Lynden-Bell}, D., \& {Pringle}, J.~E. 1974, \mnras, 168, 603,
  \dodoi{10.1093/mnras/168.3.603}

\bibitem[{{Manara} {et~al.}(2014){Manara}, {Testi}, {Natta}, {Rosotti},
  {Benisty}, {Ercolano}, \& {Ricci}}]{mana14}
{Manara}, C.~F., {Testi}, L., {Natta}, A., {et~al.} 2014, \aap, 568, A18,
  \dodoi{10.1051/0004-6361/201323318}

\bibitem[{{McMullin} {et~al.}(2007){McMullin}, {Waters}, {Schiebel}, {Young},
  \& {Golap}}]{mcmu07}
{McMullin}, J.~P., {Waters}, B., {Schiebel}, D., {Young}, W., \& {Golap}, K.
  2007, in Astronomical Society of the Pacific Conference Series, Vol. 376,
  Astronomical Data Analysis Software and Systems XVI, ed. R.~A. {Shaw},
  F.~{Hill}, \& D.~J. {Bell}, 127

\bibitem[{{Mer{\'\i}n} {et~al.}(2010){Mer{\'\i}n}, {Brown}, {Oliveira},
  {Herczeg}, {van Dishoeck}, {Bottinelli}, {Evans}, {Cieza}, {Spezzi},
  {Alcal{\'a}}, {Harvey}, {Blake}, {Bayo}, {Geers}, {Lahuis}, {Prusti},
  {Augereau}, {Olofsson}, {Walter}, \& {Chiu}}]{meri2010a}
{Mer{\'\i}n}, B., {Brown}, J.~M., {Oliveira}, I., {et~al.} 2010, \apj, 718,
  1200, \dodoi{10.1088/0004-637X/718/2/1200}

\bibitem[{{Miyake} \& {Nakagawa}(1993)}]{miya1993}
{Miyake}, K., \& {Nakagawa}, Y. 1993, \icarus, 106, 20,
  \dodoi{10.1006/icar.1993.1156}

\bibitem[{{Monet} {et~al.}(2003){Monet}, {Levine}, {Canzian}, {Ables}, {Bird},
  {Dahn}, {Guetter}, {Harris}, {Henden}, {Leggett}, {Levison}, {Luginbuhl},
  {Martini}, {Monet}, {Munn}, {Pier}, {Rhodes}, {Riepe}, {Sell}, {Stone},
  {Vrba}, {Walker}, {Westerhout}, {Brucato}, {Reid}, {Schoening}, {Hartley},
  {Read}, \& {Tritton}}]{mone2003}
{Monet}, D.~G., {Levine}, S.~E., {Canzian}, B., {et~al.} 2003, \aj, 125, 984,
  \dodoi{10.1086/345888}

\bibitem[{{Najita} {et~al.}(2015){Najita}, {Andrews}, \& {Muzerolle}}]{naji15}
{Najita}, J.~R., {Andrews}, S.~M., \& {Muzerolle}, J. 2015, \mnras, 450, 3559,
  \dodoi{10.1093/mnras/stv839}

\bibitem[{{Okuzumi} {et~al.}(2012){Okuzumi}, {Tanaka}, {Kobayashi}, \&
  {Wada}}]{okuz2012}
{Okuzumi}, S., {Tanaka}, H., {Kobayashi}, H., \& {Wada}, K. 2012, \apj, 752,
  106, \dodoi{10.1088/0004-637X/752/2/106}

\bibitem[{{Owen} \& {Kollmeier}(2019)}]{owen2019}
{Owen}, J.~E., \& {Kollmeier}, J.~A. 2019, \mnras, 487, 3702,
  \dodoi{10.1093/mnras/stz1591}

\bibitem[{{P{\'e}rez} {et~al.}(2012){P{\'e}rez}, {Carpenter}, {Chand ler},
  {Isella}, {Andrews}, {Ricci}, {Calvet}, {Corder}, {Deller}, {Dullemond},
  {Greaves}, {Harris}, {Henning}, {Kwon}, {Lazio}, {Linz}, {Mundy}, {Sargent},
  {Storm}, {Testi}, \& {Wilner}}]{pere2012}
{P{\'e}rez}, L.~M., {Carpenter}, J.~M., {Chand ler}, C.~J., {et~al.} 2012,
  \apjl, 760, L17, \dodoi{10.1088/2041-8205/760/1/L17}

\bibitem[{{Pinilla} {et~al.}(2014){Pinilla}, {Benisty}, {Birnstiel}, {Ricci},
  {Isella}, {Natta}, {Dullemond}, {Quiroga-Nu{\~n}ez}, {Henning}, \&
  {Testi}}]{pini2014a}
{Pinilla}, P., {Benisty}, M., {Birnstiel}, T., {et~al.} 2014, \aap, 564, A51,
  \dodoi{10.1051/0004-6361/201323322}

\bibitem[{{Pinilla} {et~al.}(2018){Pinilla}, {Tazzari}, {Pascucci}, {Youdin},
  {Garufi}, {Manara}, {Testi}, {van der Plas}, {Barenfeld}, {Canovas}, {Cox},
  {Hendler}, {P{\'e}rez}, \& {van der Marel}}]{pini18a}
{Pinilla}, P., {Tazzari}, M., {Pascucci}, I., {et~al.} 2018, \apj, 859, 32,
  \dodoi{10.3847/1538-4357/aabf94}

\bibitem[{{Rau} \& {Cornwell}(2011)}]{rau11}
{Rau}, U., \& {Cornwell}, T.~J. 2011, \aap, 532, A71,
  \dodoi{10.1051/0004-6361/201117104}

\bibitem[{{Rice} {et~al.}(2006){Rice}, {Armitage}, {Wood}, \&
  {Lodato}}]{rice06}
{Rice}, W.~K.~M., {Armitage}, P.~J., {Wood}, K., \& {Lodato}, G. 2006, \mnras,
  373, 1619, \dodoi{10.1111/j.1365-2966.2006.11113.x}

\bibitem[{{Riviere-Marichalar} {et~al.}(2016){Riviere-Marichalar},
  {Mer{\'\i}n}, {Kamp}, {Eiroa}, \& {Montesinos}}]{rivi2016a}
{Riviere-Marichalar}, P., {Mer{\'\i}n}, B., {Kamp}, I., {Eiroa}, C., \&
  {Montesinos}, B. 2016, \aap, 594, A59, \dodoi{10.1051/0004-6361/201527829}

\bibitem[{{Steinpilz} {et~al.}(2019){Steinpilz}, {Teiser}, \&
  {Wurm}}]{stei2019}
{Steinpilz}, T., {Teiser}, J., \& {Wurm}, G. 2019, \apj, 874, 60,
  \dodoi{10.3847/1538-4357/ab07bb}

\bibitem[{{Tazzari} {et~al.}(2017){Tazzari}, {Testi}, {Natta}, {Ansdell},
  {Carpenter}, {Guidi}, {Hogerheijde}, {Manara}, {Miotello}, {van der Marel},
  {van Dishoeck}, \& {Williams}}]{tazz2017a}
{Tazzari}, M., {Testi}, L., {Natta}, A., {et~al.} 2017, \aap, 606, A88,
  \dodoi{10.1051/0004-6361/201730890}

\bibitem[{{Testi} {et~al.}(2003){Testi}, {Natta}, {Shepherd}, \&
  {Wilner}}]{test2003}
{Testi}, L., {Natta}, A., {Shepherd}, D.~S., \& {Wilner}, D.~J. 2003, \aap,
  403, 323, \dodoi{10.1051/0004-6361:20030362}

\bibitem[{{Testi} {et~al.}(2014){Testi}, {Birnstiel}, {Ricci}, {Andrews},
  {Blum}, {Carpenter}, {Dominik}, {Isella}, {Natta}, {Williams}, \&
  {Wilner}}]{test2014}
{Testi}, L., {Birnstiel}, T., {Ricci}, L., {et~al.} 2014, in Protostars and
  Planets VI, ed. H.~{Beuther}, R.~S. {Klessen}, C.~P. {Dullemond}, \&
  T.~{Henning}, 339, \dodoi{10.2458/azu_uapress_9780816531240-ch015}

\bibitem[{{Trapman} {et~al.}(2020){Trapman}, {Ansdell}, {Hogerheijde},
  {Facchini}, {Manara}, {Miotello}, {Williams}, \& {Bruderer}}]{trap2020a}
{Trapman}, L., {Ansdell}, M., {Hogerheijde}, M.~R., {et~al.} 2020, \aap, 638,
  A38, \dodoi{10.1051/0004-6361/201834537}

\bibitem[{{van der Marel} {et~al.}(2016){van der Marel}, {Cazzoletti},
  {Pinilla}, \& {Garufi}}]{vandermarel+16}
{van der Marel}, N., {Cazzoletti}, P., {Pinilla}, P., \& {Garufi}, A. 2016,
  \apj, 832, 178, \dodoi{10.3847/0004-637X/832/2/178}

\bibitem[{{van der Marel} {et~al.}(2015){van der Marel}, {van Dishoeck},
  {Bruderer}, {P{\'e}rez}, \& {Isella}}]{vandermarel+15b}
{van der Marel}, N., {van Dishoeck}, E.~F., {Bruderer}, S., {P{\'e}rez}, L., \&
  {Isella}, A. 2015, \aap, 579, A106, \dodoi{10.1051/0004-6361/201525658}

\bibitem[{{van der Marel} {et~al.}(2013){van der Marel}, {van Dishoeck},
  {Bruderer}, {Birnstiel}, {Pinilla}, {Dullemond}, {van Kempen}, {Schmalzl},
  {Brown}, {Herczeg}, {Mathews}, \& {Geers}}]{vandermarel13}
{van der Marel}, N., {van Dishoeck}, E.~F., {Bruderer}, S., {et~al.} 2013,
  Science, 340, 1199, \dodoi{10.1126/science.1236770}

\bibitem[{{van der Marel} {et~al.}(2018){van der Marel}, {Williams}, {Ansdell},
  {Manara}, {Miotello}, {Tazzari}, {Testi}, {Hogerheijde}, {Bruderer}, {van
  Terwisga}, \& {van Dishoeck}}]{vandermarel18a}
{van der Marel}, N., {Williams}, J.~P., {Ansdell}, M., {et~al.} 2018, \apj,
  854, 177, \dodoi{10.3847/1538-4357/aaaa6b}

\bibitem[{{Villenave} {et~al.}(2019){Villenave}, {Benisty}, {Dent},
  {M{\'e}nard}, {Garufi}, {Ginski}, {Pinilla}, {Pinte}, {Williams}, {de Boer},
  {Morino}, {Fukagawa}, {Dominik}, {Flock}, {Henning}, {Juh{\'a}sz}, {Keppler},
  {Muro-Arena}, {Olofsson}, {P{\'e}rez}, {van der Plas}, {Zurlo}, {Carle},
  {Feautrier}, {Pavlov}, {Pragt}, {Ramos}, {Sauvage}, {Stadler}, \&
  {Weber}}]{vill2019a}
{Villenave}, M., {Benisty}, M., {Dent}, W.~R.~F., {et~al.} 2019, \aap, 624, A7,
  \dodoi{10.1051/0004-6361/201834800}

\bibitem[{{Wada} {et~al.}(2009){Wada}, {Tanaka}, {Suyama}, {Kimura}, \&
  {Yamamoto}}]{wada2009}
{Wada}, K., {Tanaka}, H., {Suyama}, T., {Kimura}, H., \& {Yamamoto}, T. 2009,
  \apj, 702, 1490, \dodoi{10.1088/0004-637X/702/2/1490}

\bibitem[{{Weidenschilling}(1977)}]{weid77}
{Weidenschilling}, S.~J. 1977, \mnras, 180, 57, \dodoi{10.1093/mnras/180.1.57}

\bibitem[{{Weidenschilling}(1984)}]{weid1984a}
---. 1984, \icarus, 60, 553, \dodoi{10.1016/0019-1035(84)90164-7}

\bibitem[{{Weidenschilling} \& {Cuzzi}(1993)}]{weid1993}
{Weidenschilling}, S.~J., \& {Cuzzi}, J.~N. 1993, in Protostars and Planets
  III, ed. E.~H. {Levy} \& J.~I. {Lunine}, 1031

\bibitem[{{Whitney} {et~al.}(2013){Whitney}, {Robitaille}, {Bjorkman}, {Dong},
  {Wolff}, {Wood}, \& {Honor}}]{whit2013}
{Whitney}, B.~A., {Robitaille}, T.~P., {Bjorkman}, J.~E., {et~al.} 2013, \apjs,
  207, 30, \dodoi{10.1088/0067-0049/207/2/30}

\bibitem[{{Wood} {et~al.}(2002){Wood}, {Wolff}, {Bjorkman}, \&
  {Whitney}}]{wood2002}
{Wood}, K., {Wolff}, M.~J., {Bjorkman}, J.~E., \& {Whitney}, B. 2002, \apj,
  564, 887, \dodoi{10.1086/324285}

\bibitem[{{Yen} {et~al.}(2018){Yen}, {Koch}, {Manara}, {Miotello}, \&
  {Testi}}]{yen2018a}
{Yen}, H.-W., {Koch}, P.~M., {Manara}, C.~F., {Miotello}, A., \& {Testi}, L.
  2018, \aap, 616, A100, \dodoi{10.1051/0004-6361/201732196}

\bibitem[{{Zhang} {et~al.}(2016){Zhang}, {Bergin}, {Blake}, {Cleeves},
  {Hogerheijde}, {Salinas}, \& {Schwarz}}]{zhan16}
{Zhang}, K., {Bergin}, E.~A., {Blake}, G.~A., {et~al.} 2016, \apjl, 818, L16,
  \dodoi{10.3847/2041-8205/818/1/L16}

\bibitem[{{Zhu} {et~al.}(2012){Zhu}, {Nelson}, {Dong}, {Espaillat}, \&
  {Hartmann}}]{zhu12}
{Zhu}, Z., {Nelson}, R.~P., {Dong}, R., {Espaillat}, C., \& {Hartmann}, L.
  2012, \apj, 755, 6, \dodoi{10.1088/0004-637X/755/1/6}

\end{thebibliography}
\bibliographystyle{aasjournal}

\end{document}